%% file: neo.tex
\documentclass{vldb}
\usepackage{graphicx}
\usepackage{balance}  % for  \balance command ON LAST PAGE  (only there!)
\usepackage{xspace}
\usepackage{url}
\usepackage{cite}
\usepackage{enumitem}
\usepackage{tabularx}

\usepackage{amssymb}
\usepackage{subcaption}
\usepackage{color}
\usepackage{array}
\usepackage{comment}
\usepackage{xcolor,listings}
\usepackage{textcomp}
\usepackage{tikz-qtree}
\usepackage{tikz}
\usepackage{booktabs}
\newsavebox{\mybox}

\usepackage{graphicx}

\makeatletter
\def\blfootnote{\xdef\@thefnmark{}\@footnotetext}
\makeatother

\newcommand{\sparagraph}[1]{\vspace{1mm}\noindent {\bf #1}}

\newcommand{\JOB}[0]{\texttt{JOB}\xspace}
\newcommand{\EJOB}[0]{\texttt{Ext-JOB}\xspace}
\newcommand{\TPC}[0]{\texttt{TPC-H}\xspace}
\newcommand{\Vertica}[0]{\texttt{Corp}\xspace}
\newcommand{\oh}[0]{\texttt{1-Hot}\xspace}
\newcommand{\hist}[0]{\texttt{Histogram}\xspace}
\newcommand{\rv}[0]{\texttt{R-Vector}\xspace}

\newcommand{\system}[0]{Neo\xspace}

\newcommand{\PG}[0]{PostgreSQL\xspace}
\newcommand{\SQLite}[0]{SQLite\xspace}
\newcommand{\MS}[0]{MS SQL Server\xspace}
\newcommand{\Ora}[0]{Oracle\xspace}

\vldbTitle{\system}
\vldbAuthors{Brandeis and MIT people}
\vldbDOI{https://doi.org/10.14778/xxxxxxx.xxxxxxx}
\vldbVolume{12}
\vldbNumber{xxx}
\vldbYear{2019}

\begin{document}

% ****************** TITLE ****************************************

\title{\system: A Learned Query Optimizer}

% possible, but not really needed or used for PVLDB:
%\subtitle{[Extended Abstract]
%\titlenote{A full version of this paper is available as\textit{Author's Guide to Preparing ACM SIG Proceedings Using \LaTeX$2_\epsilon$\ and BibTeX} at \texttt{www.acm.org/eaddress.htm}}}

% ****************** AUTHORS **************************************


\numberofauthors{1} %  in this sample file, there are a *total*

\author{
  Ryan Marcus$^{1}$, 
  Parimarjan Negi$^{2}$, 
  Hongzi Mao$^{2}$, 
  Chi Zhang$^{1}$,\\
  Mohammad Alizadeh$^{2}$, 
  Tim Kraska$^{2}$, 
  Olga Papaemmanouil$^{1}$, 
  Nesime Tatbul$^{23}$ \\ \smallskip
  \affaddr{$^1$Brandeis University \quad $^2$MIT \quad $^3$Intel Labs}
\\\
% 1st. author
%\alignauthor
%Ben Trovato\titlenote{Dr.~Trovato insisted his name be first.}\\
%       \affaddr{Institute for Clarity in Documentation}\\
%       \affaddr{1932 Wallamaloo Lane}\\
%       \affaddr{Wallamaloo, New Zealand}\\
%       \email{trovato@corporation.com}
%% 2nd. author
%\alignauthor
%G.K.M. Tobin\titlenote{The secretary disavows
%any knowledge of this author's actions.}\\
%       \affaddr{Institute for Clarity in Documentation}\\
%       \affaddr{P.O. Box 1212}\\
%       \affaddr{Dublin, Ohio 43017-6221}\\
%       \email{webmaster@marysville-ohio.com}
%% 3rd. author
%\alignauthor Lars Th{\Large{\sf{\o}}}rv{$\ddot{\mbox{a}}$}ld\titlenote{This author is the
%one who did all the really hard work.}\\
%       \affaddr{The Th{\large{\sf{\o}}}rv{$\ddot{\mbox{a}}$}ld Group}\\
%       \affaddr{1 Th{\large{\sf{\o}}}rv{$\ddot{\mbox{a}}$}ld Circle}\\
%       \affaddr{Hekla, Iceland}\\
%       \email{larst@affiliation.org}
%\and  % use '\and' if you need 'another row' of author names
%% 4th. author
%\alignauthor Lawrence P. Leipuner\\
%       \affaddr{Brookhaven Laboratories}\\
%       \affaddr{Brookhaven National Lab}\\
%       \affaddr{P.O. Box 5000}\\
%       \email{lleipuner@researchlabs.org}
%% 5th. author
%\alignauthor Sean Fogarty\\
%       \affaddr{NASA Ames Research Center}\\
%       \affaddr{Moffett Field}\\
%       \affaddr{California 94035}\\
%       \email{fogartys@amesres.org}
%% 6th. author
%\alignauthor Charles Palmer\\
%       \affaddr{Palmer Research Laboratories}\\
%       \affaddr{8600 Datapoint Drive}\\
%       \affaddr{San Antonio, Texas 78229}\\
%       \email{cpalmer@prl.com}
}
% There's nothing stopping you putting the seventh, eighth, etc.
% author on the opening page (as the 'third row') but we ask,
% for aesthetic reasons that you place these 'additional authors'
% in the \additional authors block, viz.
%\additionalauthors{Additional authors: John Smith (The Th{\o}rv\"{a}ld Group, {\texttt{jsmith@affiliation.org}}), Julius P.~Kumquat
%(The \raggedright{Kumquat} Consortium, {\small \texttt{jpkumquat@consortium.net}}), and Ahmet Sacan (Drexel University, {\small \texttt{ahmetdevel@gmail.com}})}
% Just remember to make sure that the TOTAL number of authors
% is the number that will appear on the first page PLUS the
% number that will appear in the \additionalauthors section.

\maketitle

\begin{abstract}

Query optimization is one of the most challenging problems in database systems. Despite the progress made over the past decades, query optimizers remain extremely complex components that require a great deal of hand-tuning for specific workloads and datasets. Motivated by this shortcoming and inspired by recent advances in applying machine learning to data management challenges, we introduce \emph{Neo} (\emph{Neural Optimizer}), a novel learning-based query optimizer that relies on deep neural networks  to generate query executions plans. 
Neo bootstraps its query optimization model from existing optimizers and continues to learn from incoming queries, building upon its successes and learning from its failures.  Furthermore, Neo naturally adapts to underlying data patterns and is robust to estimation errors. Experimental results demonstrate that Neo, even when bootstrapped from a simple optimizer like PostgreSQL,  can learn a model that offers similar performance to state-of-the-art commercial optimizers, and in some cases even surpass them. 
%\blfootnote{$^\dagger$ Listed alphabetically.}
\end{abstract}

\input{introduction}
\input{system_model}
\input{lo}

\input{notation}

\input{lo_features}

\input{lo_value_network}
\input{lo_search}

\input{pari_features}
\input{experiments.tex}
\input{related}
\input{conclusions}
\bibliographystyle{abbrv}
\bibliography{neo}  % vldb_sample.bib is the name of the Bibliography in this case

%APPENDIX is optional.
% ****************** APPENDIX **************************************
% Example of an appendix; typically would start on a new page
%pagebreak
\appendix
\input{apx_model}
%\begin{appendix}
%\end{appendix}

\end{document}

%% file: introduction.tex
\section{Introduction}
\label{sec:intro}

%Since the countless success stories of machine learning,
In the face of a never-ending deluge of machine learning success stories,
every database researcher has likely wondered if it is possible to \emph{learn} a query optimizer. 
Query optimizers are key to achieving good performance in database systems, and can  speed up the execution time of queries by orders of magnitude.
However, building a good optimizer today takes thousands of person-engineering-hours, and is an art only a few experts fully master. 
Even worse, query optimizers need to be tediously maintained, especially as the system's execution and storage engines evolve. 
As a result, none of the freely available open-source query optimizers come close to the performance of the commercial optimizers offered by IBM, Oracle, or Microsoft.
% none can be singular or plural. here as not any I like plural better
%, as neither the expertise nor the engineering-power are available. 

%As optimizers are essentially built from tens to hundreds of carefully hand-tuned heuristics, they are an obvious target for machine learning and at no surprise there have been many attempts to improve query optimizers through learning over the last decades. \ma{This transition seems kind of repetitive. Maybe just start with: There have been many attempts to improve...}

Due to the heuristic-based nature of query optimization, there have been many attempts to improve query optimizers through learning over the last several decades.
For example, almost two decades ago, Leo, DB2's {LE}arning Optimizer, was proposed~\cite{leo}. 
Leo learns from its mistakes by adjusting its cardinality estimations over time. 
However, Leo still requires a human-engineered cost model, a hand-picked search strategy, and a lot of developer-tuned heuristics, which take years to develop and perfect.
Furthermore, Leo only learns better cardinality estimations, but not new optimization strategies (e.g., how to account for uncertainty in cardinality estimates, operator selection, etc.).

%Furthermore it only learns better cardinality estimations, but not new optimization strategies (e.g., when to pick a more robust plan vs. a highly-optimized but variable plan \ma{will people understand what we mean by robust vs. highly optimized plans? something like `how to account for uncertainty in cardinality estimates' might be easier to follow}).

More  recently, the database community  has started to explore how neural networks can be used  to improve query optimizers~\cite{dbml, cidr_dlqo}. For example, DQ~\cite{sanjay_wat} and ReJOIN~\cite{rejoin} use reinforcement learning combined with a human-engineered cost model to automatically learn search strategies to explore the space of possible join orderings. While these papers show that learned search strategies can outperform conventional heuristics on the provided cost model, they do not show a consistent or significant impact on actual query performance. Moreover, they still rely on the optimizer's heuristics  for cardinality estimation,  physical operator selection, and estimating the cost of candidate execution plan. 

%However, the work still relies on a hand-tuned cost-model and heuristics for cardinality estimation and physical operator selection, the model can take up to days to train, and in the end it does not show if the search strategy actually has a significant impact on the overall query performance \ma{It might be good to give some credit while bashing these papers, e.g., While these papers show that learned search strategies can outperform conventional heuristics on the provided cost model, they do not show that there is a significant impact on actual query performance}. 
Other approaches demonstrate how machine learning can be used to achieve better cardinality estimates~\cite{deep_card_est2, quicksel, qo_state_rep}. However, none demonstrate that their improved cardinality estimations \emph{actually lead to better query plans.} It is relatively easy to improve the average error of a cardinality estimation, but much harder to improve estimations for the cases that actually improve query plans~\cite{job2}. 
Furthermore, cardinality estimation is only one component of an optimizer. Unlike join order selection, identifying join operators (e.g., hash join, merge join) and selecting indexes cannot be (entirely) reduced to cardinality estimation. 
Finally, SkinnerDB showed that adaptive query processing strategies can benefit from reinforcement learning, but requires a specialized query execution engine, and cannot benefit from operator pipelining or other advanced parallelism models~\cite{skinnerdb}.
%it not really even tries to build a query optimizers. 
%\ma{Is re-optimization a standard term in this context? I don't know what it means exactly. The last sentence could use some tinkering. Do we mean that they didn't evaluate it?} 

In other words, none of the recent machine-learning-based  approaches come close to learning an \emph{entire} optimizer, nor do they show how their techniques can achieve state-of-the-art performance (to the best of our knowledge, none of these approaches compare with a commercial optimizer).
%Instead rather weak baselines like the PostgreSQL optimizer are used. 
{Showing that an entire optimizer can be learned remains an important milestone and has far reaching implications. } %\ma{Stylistically, I'd prefer not having entirely bold sentences. It feels like we're shouting at the reader but maybe its just me. :)}
Most importantly, if a learned query optimizer could achieve performance comparable to commercial systems after a short amount of training, the amount of human development time to create a new query optimizer will be significantly reduced. 
This, in turn, will make good optimizers available to a much broader range of systems, and could improve the performance of thousands of applications that use open-source databases today. 
Furthermore, it could change the way query optimizers are built, replacing an expensive stable of heuristics with a more holistic optimization problem.
This should result in better maintainability, as well as lead to optimizers that will truly learn from their mistakes and adjust their entire strategy for a particular customer instance to achieve {\em instance optimality}~\cite{sagedb}.

%In this work, we make a big step forwards to reaching the milestone of building a very first entirely learned optimizer with state-of-the-art performance. 
In this work, we take a significant step towards the milestone of building an {\em end-to-end} learned optimizer with state-of-the-art performance. 
To the best of our knowledge, \emph{this work is the first to show that an entire query optimizer can be learned.}  Our learned optimizer is able to achieve similar performance to state-of-the-art commercial optimizers, e.g.,  Oracle and Microsoft, and sometimes even surpass them.
This required overcoming several key challenges, from
capturing query semantics as vectors,
processing tree-based query plan structures,
designing a search strategy,
incorporating physical operator and index selection,
replacing human-engineered cost models with a neural network,
adopting reinforcement learning techniques to continuously improve,
and significantly shorting the training time for a given database. 
All these techniques were integrated into the first end-to-end learned query optimizer, called \emph{Neo} (\emph{Neural Optimizer}). 

\system learns to make decisions about join ordering, physical operator selection, and index selection. However, we have not yet reached the milestone of learning these tasks from scratch.
First, \system still requires a-priori knowledge about all possible query rewrite rules (this guarantees semantic correctness and the number of rules are usually small). 
Second, we restrict \system to project-select-equijoin-aggregate-queries (though, our framework is general and can easily be extended). 
Third, our optimizer does not yet generalize from one database to another, as our features are specific to a set of tables\,---\,however, \system \emph{does} generalize to unseen queries, which can contain any number of known tables.
%We note that the number of rewrite rules is relatively small, and the restrictions to project-select-equijoin-aggregate-queries is not fundamental (for example, we can already support subqueries).
%That we still need a traditional optimizer and can not even bootstrap from our own optimizer, is a downside of our approach. We therefore have to be very careful how we write it as one might say, oh they still need Postgres and is not yet End-to-End. 
Fourth, \system requires a traditional (weaker) query optimizer to bootstrap its learning process.
As proposed in~\cite{cidr_dlqo}, we use the traditional query optimizer as a source of expert demonstration, which \system uses to bootstrap its initial policy. This technique, referred to as ``learning from demonstration''~\cite{demonstration,dqfd,pretrain_demonstration,lift} significantly speeds up the learning process, reducing training time from days/weeks to just a few hours. The query optimizer used to bootstrap \system can be much weaker in performance and, after an initial training period, \system surpasses its performance and it is no longer needed. 
For this work, we use the \PG optimizer, but any  traditional (open source) optimizer can be used.

%%%%%%%%%%%%%%%%%%%%%%%%%%%%%
%% 
% Fourth, \system requires a traditional (weaker) query optimizer to bootstrap its learning process.
% As proposed in~\cite{cidr_dlqo}, we use the traditional query optimizer as a source of expert demonstration, which \system uses to bootstrap its initial policy. This technique, referred to as ``learning from demonstration'', is a standard technique~\cite{demonstration,dqfd,pretrain_demonstration} to significant speed-up the learning process and is already used in other ``ML for Systems'' approaches~\cite{lift}.
% However, the query optimizer used to bootstrap \system can be much weaker in performance and, after an initial training period, is no longer needed. 
% For this work, we use the \PG optimizer, but any  traditional (open source) optimizer can be used.

% Learning from demonstration is critical, as it reduces training time from days/weeks to just a few hours, ultimately enabling \system to quickly surpass the \PG optimizer in performance on a variety of workloads.
% Moreover, it often even outperforms commercial optimizers on their own query execution engines even when it is boostrapped using the PostgreSQL optimizer. 

%%%%%%%%%%%%%%%

Our results show that \system outperforms commercial optimizers on their own query execution engines, even when it is boostrapped using the PostgreSQL optimizer. Interestingly, \system learns to automatically adjust to changes in the accuracy of cardinality predictions (e.g., it picks more robust plans if the cardinality estimates are less precise). Further, it can be tuned depending on the customer preferences (e.g., increase worst-case performance vs. relative performance)\,---\,adjustments which are  hard to achieve with traditional techniques.

We argue that \system \emph{represents a step forward in building an entirely learned optimizer.} 
Moreover, \system can already be used, in its current form, to improve the performance of thousands of applications which rely on PostgreSQL and other open-source database systems (e.g. SQLite).
We hope that \system inspires many other database researchers to experiment with combining query optimizers and learned systems in new ways, similar to how AlexNet~\cite{alexnet} changed the way image classifiers were built.

In summary, we make the following contributions:
\begin{itemize}[nosep]
\item We propose \system\,---\,an end-to-end learning approach to query optimization, including join ordering, index selection, and physical operator selection. 
%\item We introduce a number of novel compon \emph{value model},  a deep neural network  designed to predict the \emph{final execution time} of a given partial execution plan
\item We show that, after training for a dataset and representative sample query workload, \system is able to generalize even over queries it has not encountered before.  
\item We evaluate different feature engineering techniques and propose a new technique, which implicitly represents correlations within the dataset.  
\item We show that, after a short amount of training, \system is able to achieve performance comparable to Oracle's and Microsoft's query optimizers on their respective database systems. 
\end{itemize}

The remainder of the paper is organized as follows: Section~\ref{sec:overview} provides an overview of \system's learning framework and system model. Section~\ref{sec:lo} describes how queries and query plans are represented by \system. Section~\ref{sec:value_network} explains \system's value network, the core learned component of our system. Section~\ref{sec:pari_features} describes row vector embeddings, an optional learned representation of the underlying database that helps \system understand correlation within the user's data. We present an experimental evaluation of \system in Section~\ref{sec:experiments}, discuss related works in Section~\ref{sec:related}, and offer concluding remarks in Section~\ref{sec:conclusions}.

%% file: system_model.tex
\begin{table*}
\centering
\begin{tabularx}{\textwidth}{lp{5cm}X}
\toprule
  & {\bf Traditional Optimizer} & {\bf Neural Optimizer (\system)} \\
\midrule
{\bf Creation} & Human developers & Demonstration, reinforcement learning {\em (Section~\ref{sec:overview})} \\
{\bf Query Representation} &  Operator tree & Feature encoding {\em (Section~\ref{sec:lo})} \\

{\bf Cost Model} & Hand-crafted model &  Learned DNN model {\em (Section~\ref{sec:value_network})}\\

{\bf Plan Space Enumeration} & Heuristics, dynamic programming & DNN-guided search strategy {\em (Section~\ref{sec:search})} \\

{\bf Cardinality Estimation} & Histograms, hand-crafted models & Histograms, learned embeddings {\em (Section~\ref{sec:pari_features})}\\
\bottomrule
\end{tabularx}
\caption{Traditional cost-based query optimizer vs. Neo}
\label{tab:compare}
\end{table*}

\section{Learning Framework Overview}
\label{sec:overview}

\begin{figure}
  \centering
  \includegraphics[width=0.45\textwidth]{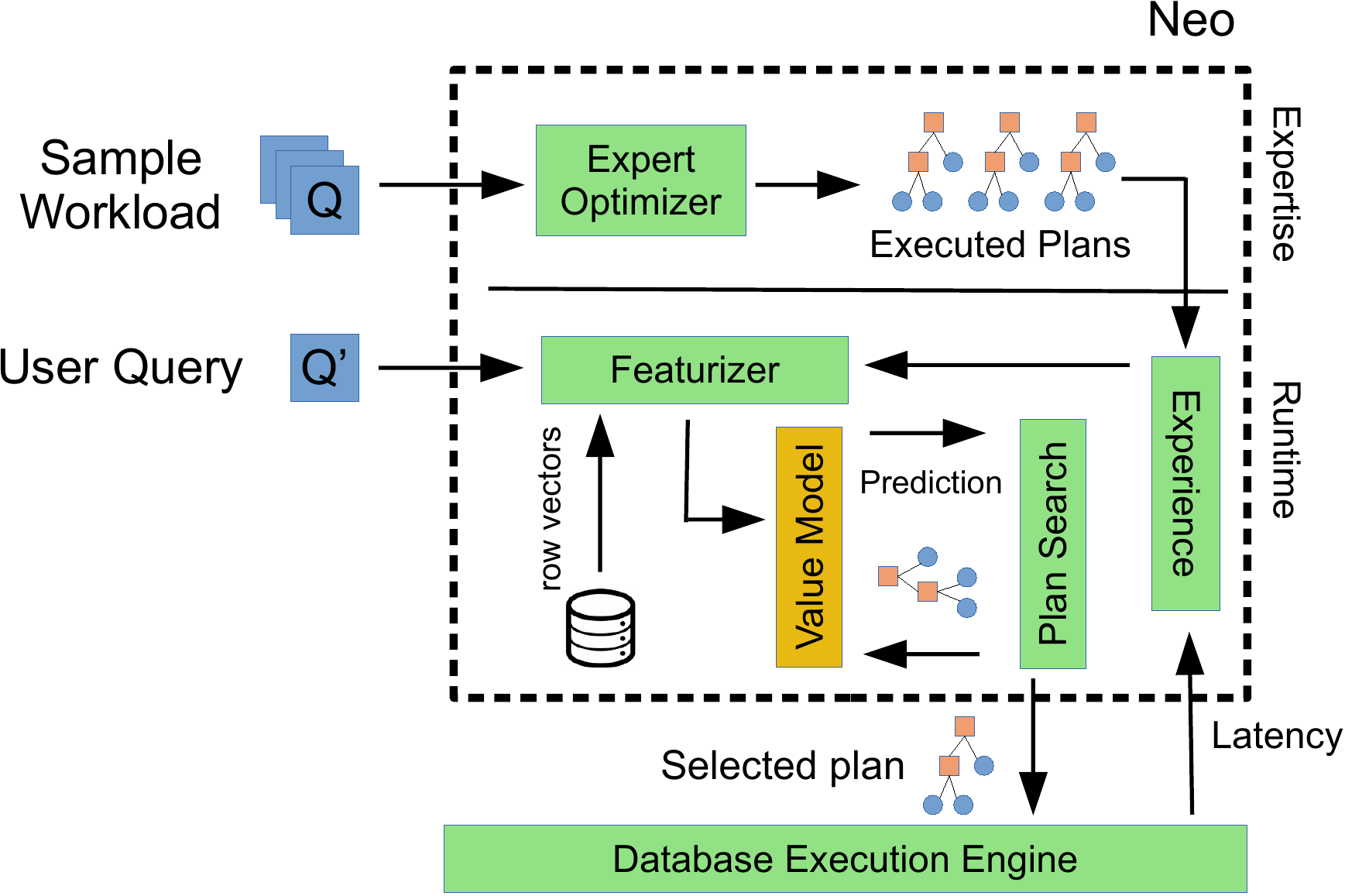}
  \caption{\system system model}
  \label{fig:system_model}
  \vspace{-5mm}
\end{figure}

What makes \system unique that it is the very first end-to-end query optimizer. 
As shown in Table~\ref{tab:compare}, it replaces every component of a traditional Selinger-style~\cite{systemr} query optimizers through machine learning models:
(i) the query representation is through features rather than an object-based query operator tree (e.g., Volcano-style~\cite{volcano} iterator tree);
(ii) the cost model is a deep neural network (DNN) model as opposed to hand-crafted equations;
(iii) the search strategy is a DNN-guided learned best-first search strategy instead of plan space enumeration or dynamic programming;
(iv) cardinality estimation is based on either histograms or a learned vector embedding scheme, combined with a learned model, instead of hand-tuned histogram-based cardinality estimation model. 
Finally, (v) Neo uses reinforcement learning and learning from demonstration to integrate these into an end-to-end query optimizer rather than relying on human engineering.
While we describe the different components in the individual sections as outlined in Table~\ref{tab:compare}, the following provides a general overview of how \system learns, as depicted in Figure~\ref{fig:system_model}.

\sparagraph{Expertise Collection} The first phase, labeled \emph{Expertise}, initially generates experience from a traditional query optimizer, as proposed in~\cite{cidr_dlqo}. 
 \system assumes the existence of an application-provided \emph{Sample Workload} consisting of queries representative of the application's total workload.  Additionally, we assume \system  has access to a simple, traditional rule- or cost-based {\em Expert Optimizer} (e.g., Selinger~\cite{systemr}, PostgreSQL~\cite{url-postgres}). This simple optimizer is treated as a black box, and is \emph{only} used to create query execution plans (QEPs) for each query in the sample workload. These QEPs, along with their latencies, are added to \system's {\em Experience} (i.e., a set of plan/latency pairs), which will be used as a starting point in the next model training phase. Note that the {\em Expert Optimizer} can be \emph{unrelated} to the underlying {\em Database Execution Engine}.

\sparagraph{Model Building} Given the collected experience, \system builds an initial {\em Value Model}. The value model is a deep neural network with an architecture designed to predict the \emph{final} execution time of a given partial or complete plan for a given query. We train the value network using the collected experience in a supervised fashion. This process involves transforming each user-submitted query into features (through the \emph{Featurizer} module) useful for a machine learning model. These features contain both query-level information (e.g., the join  graph, predicated attributes, etc.) and plan-level information (e.g., selected join order, access paths, etc.). \system can work with a number of different featurization techniques, ranging from simple one-hot encodings (Section \ref{sec:features}) to more complex embeddings (Section \ref{sec:pari_features}). \system's value network uses tree convolution~\cite{tree_conv} to process the tree-structured QEPs (Section~\ref{sec:tree_conv}).

% Given the feature vectors of each query and its execution time, the value model is trained to predict the \emph{final} query execution time of each partial and full QEP for a given query.

% At its core, the value model uses tree convolution~\cite{tree_conv} in order to take the tree-structure of QEPs  as input (\S\ref{sec:tree_conv}). 
% Upon arrival of a user-query, \system uses the built model to generate a QEP for the new query. Specifically, the new query is transformed into a \emph{global feature vector} and passed as input to the value model. \system can work with a number of different featurization techniques, ranging from simple 1-hot encodings to more complex embeddings (\S\ref{sec:pari_features}). %This global feature vector serves as the first of two inputs to the value model.

\sparagraph{Plan Search} Once query-level information has been encoded, \system uses the value model to search over the space of QEPs (i.e., selection of join orderings, join operators, and indexes) and discover the plan with the minimum predicted execution time (i.e., value).  Since the space of all execution plans for a particular query is far too large to exhaustively search, \system performs a best-first search of the space, using the value model as a heuristic. A complete plan created by \system, which includes a join ordering, join operators (e.g. hash, merge, loop), and access paths (e.g., index scan, table scan) is sent to the underlying execution engine, which is responsible for applying semantically-valid query rewrite rules (e.g., inserting necessary hash and sort operations) and executing the final plan. This ensures that every execution plan generated by \system computes the correct result.
% Note that, because \system's value model attempts to predict the best-possible \emph{final} latency of a partial query plan,  dynamic programming often used for query optimization~\cite{systemr} (whose runtime increases exponentially with the  number of joins in a query) are not required: searching for the best plan runs in time linear to the number of joins in the query.
The plan search is discussed in detail in Section~\ref{sec:search}.

%\ma{I don't fully understand the point we're trying to make here. maybe it's ok if we explain it later, but perhaps a few sentences about the alternative approach (that is exponential) could help}. 

\sparagraph{Model Refinement} As new queries are optimized through \system, the model is iteratively improved and custom-tailored to the underlying database and execution engine. This is achieved by incorporating  newly collected experience regarding each query's QEP and performance. Specifically, once a QEP is chosen for a particular query, it is sent to the underlying execution engine, which processes the query and returns the result to the user. Additionally, \system records the final execution latency of the QEP, adding the plan/latency pair to its \emph{Experience}. Then, \system retrains the value model based on this experience, iteratively improving its estimates.  
% The value network is trained in a supervised fashion: for each observed QEP, the value network is trained to map each subplan of the observed QEP to the best final latency observed so far from a QEP with that subplan.

\sparagraph{Discussion} This process -- featurizing, searching, and refining -- is repeated for each query sent by the user. \system's architecture is designed to create a corrective feedback loop: when \system's learned cost model guides \system to a query plan that \system predicts will perform well, but then the resulting latency is high, \system's cost model learns to predict a higher cost for the poorly-performing plan. Thus, \system is less likely to choose plans with similar properties to the poorly-performing plan in the future. As a result, \system's cost model becomes more accurate, effectively \emph{learning from its mistakes.}

\system's architecture, of using a learned cost model to guide a search through a large and complex space, is inspired by AlphaGo~\cite{alphago}, a reinforcement learning system developed to play the game of Go. At a high level, for each move in a game of Go, AlphaGo uses a neural network to evaluate the desirability of each potential move, and uses a search routine to find a sequence of moves that is most likely to lead to a winning position. Similarly, \system uses a neural network to evaluate the desirability of partial query plans, and uses a search function to find a complete query plan that is likely to lead to lower latency.%\footnote{AlphaGo's search strategy is significantly more complex than \system's, as AlphaGo must account for an adversary. \ma{I'm not sure this is relevant. I think AlphaGo is more complex because it has to consider a much large space of moves and also search much deeper. So for example, AlphaGo can't consider all moves at a particular step in the way that Neo does. Instead it uses a policy network, in addition to the value network, to pruce the search space.}} 

Both AlphaGo and \system additionally bootstrap their cost models from experts. AlphaGo bootstraps from a dataset of Go games played by expert humans, and \system bootstraps from a dataset of query execution plans built by a traditional query optimizer (which was designed by human experts). The reason for this bootstrapping is because of reinforcement learning's inherent \emph{sample inefficiency}~\cite{dqfd,demonstration}: without bootstrapping, reinforcement learning algorithms like \system or AlphaGo may require millions of iterations~\cite{dqn} before even becoming competitive with human experts. Bootstrapping from an expert source (i.e., learning from demonstration) intuitively mirrors how young children acquire language and learn to walk by imitating nearby adults (experts), and has been shown to drastically reduce the number of iterations required to learn a good policy~\cite{dqfd,lift}. Decreasing sample inefficiency is especially critical for database management systems: each iteration requires a query execution, and users are unlikely to be willing to execute millions of queries before achieving performance on-par with current query optimizers. Worse yet, executing a poor query execution plan takes \emph{longer} than executing a good execution plan, so the initial iterations would take an infeasible amount of time to complete~\cite{cidr_dlqo}.

Thus, \system can be viewed as a learning-from-demonstration reinforcement learning system similar to AlphaGo -- there are, however, many differences between AlphaGo and \system.
%(\system is significantly more complicated than ``AlphaGo for query optimization'').
First, because of the grid-like nature of the Go board, AlphaGo can trivially represent the board as an image and use image convolution, possibly the most well-studied and highly-optimized neural network primitive~\cite{deep_survey, deep_learning}, in order to predict the desirability of a board state. On the other hand, query execution plans have a tree structure, and cannot be trivially represented as images, nor can image convolution be easily applied. Second, in Go, the board represents all the information relevant to a particular move, and can be represented using less than a kilobyte of storage. In query optimization, the data in the user's database is highly relevant to the performance of query execution plans, and is generally \emph{much} larger than a kilobyte (it is not possible to simply feed a user's entire database into a neural network). Third, AlphaGo has a single, unambiguous goal: defeat its opponent and reach a winning game state. \system, on the other hand, needs to take the user's preferences into account, e.g. should \system optimize for average-case or worst-cast latency?

The remainder of this paper describes our solutions to these problems in detail, starting with the notation and encoding of the query plans.

%% file: lo.tex
\section{Query Featurization}
\label{sec:lo}

In this section, we describe how query plans are represented as vectors, starting with  some necessary notation. %Then, we describe the query-specific features extracted and used by \system, along with how each node in an execution plan is represented.

%%% Local Variables:
%%% mode: latex
%%% TeX-master: "main"
%%% End:

%% file: notation.tex
\subsection{Notation}

\begin{figure}
    \centering
    \includegraphics[width=0.24\textwidth]{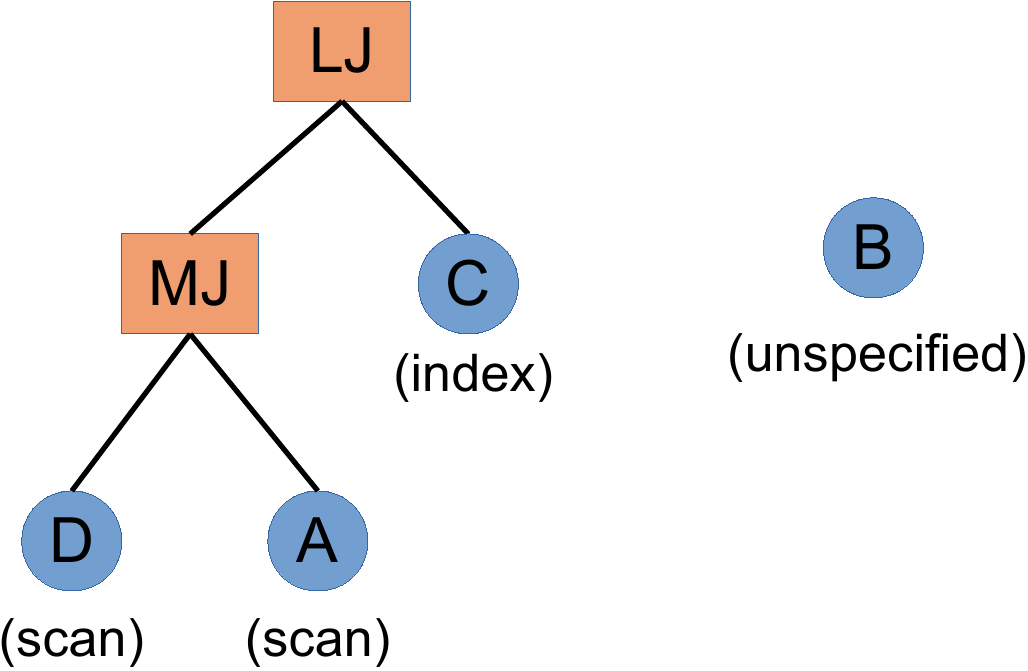}
    \caption{Partial query plan}
    \label{fig:partial_plan}
\end{figure}

For a given query $q$, we define the set of base relations used in $q$ as $R(q)$. A partial execution plan $P$ for a query $q$ (denoted $Q(P) = q$) is a forest of trees representing an execution plan that is still being built. Each internal (non-leaf) tree node is a join operator $\bowtie_i \in J$, where $J$ is the set of possible join operators (e.g., hash join $\bowtie_H$, merge join $\bowtie_M$, loop join $\bowtie_L$) and each leaf tree node is either a table scan, an index scan, or an unspecified scan over a relation $r \in R(q)$, denoted $T(r)$, $I(r)$, and $U(r)$ respectively.\footnote{\system can trivially handle additional scan types, e.g., bitmap scans.} An unspecified scan is a scan that has not been assigned as either a table or an index scan yet. For example, the partial query execution plan depicted in Figure~\ref{fig:partial_plan} is denoted as:
\begin{equation*}
\left[(T(D) \bowtie_M T(A)) \bowtie_L I(C)\right], \quad \left[U(B)\right]
\end{equation*}

Here, the type of scan for $B$ is still unspecified, as is the join that will eventually link $B$ with the rest of the plan, but the plan specifies a table scan of table $D$ and $A$, which feed into a merge join, whose result will then be joined using a loop join with $C$. 

A \emph{complete} execution plan is a  plan with only a single tree and with no unspecified scans; all decisions on how the plan should be executed have been made.
We say that one execution plan $P_i$ is a \emph{subplan} of another execution plan $P_j$, written $P_i \subset P_j$, if $P_j$ could be constructed from $P_i$ by (1) replacing unspecified scans with index or table scans, and (2) combining subtrees in $P_i$ with a join operator. %Intuitively, $P_i$ is a subplan of $P_j$ if $P_j$ can be built from $P_i$. \MA{repetitive}
%For example, $P_1$ is a subplan of $P_2$.

%%% Local Variables:
%%% mode: latex
%%% TeX-master: "main"
%%% End:

%% file: lo_features.tex
\subsection{Encodings}

In order to train a neural network to predict the final latency of  partial or complete query execution plans (QEPs), we require two encodings: first, a {\em query encoding}, which encodes information regarding the query, but is independent of the query plan. For example, the involved tables and predicates fall into this category. Second, we require a {\em plan encoding}, which represents the partial execution plan. 

%Old text
%In order to train a neural network to predict the \emph{final} latency of  partial query execution plans (QEPs), we must first represent partial query execution plans in a way that is amenable to the neural network, i.e., to encode partial QEPs as feature vectors. It is critical that this representation contains sufficient information about a given partial plan to enable the network to learn patterns relating query execution plans to final latencies.
%We encode partial execution plans in two parts: (1) \emph{query-level information}, which is information that is related to the query, and is thus shared between all partial execution plans for that query, and (2) \emph{node-level information}, which is information about a particular tree node in the partial execution plan.

\sparagraph{Query Encoding} The representation of query-dependent but plan-independent information in \system is similar to the representations used in previous work~\cite{sanjay_wat, rejoin}, and consists of two components. The first component encodes the joins performed by the query, which can be represented as an adjacency matrix of the join graph, e.g. in Figure~\ref{fig:features}, the 1 in the first row, third column corresponds to the join predicate connecting A and C. Both the row and column corresponding to the relation $E$ are empty, because $E$ is not involved in the example query.
Here, we assume that at most one foreign key between each relation exists. However, the representation can easily be extended to include more than one potential foreign key (e.g., by using the index of the relevant key instead of the constant value ``1'', or by adding additional columns for each foreign key). 
Furthermore, since this matrix is symmetrical, we choose only to encode the upper triangular portion, colored in red in Figure~\ref{fig:features}.
Note that the join graph does not specify the order of joins.

The second component of the query encoding is the column predicate vector. In \system, we currently support three increasingly powerful variants, with varying levels of precomputation requirements:

\begin{figure}
    \centering
    \includegraphics[width=0.4\textwidth]{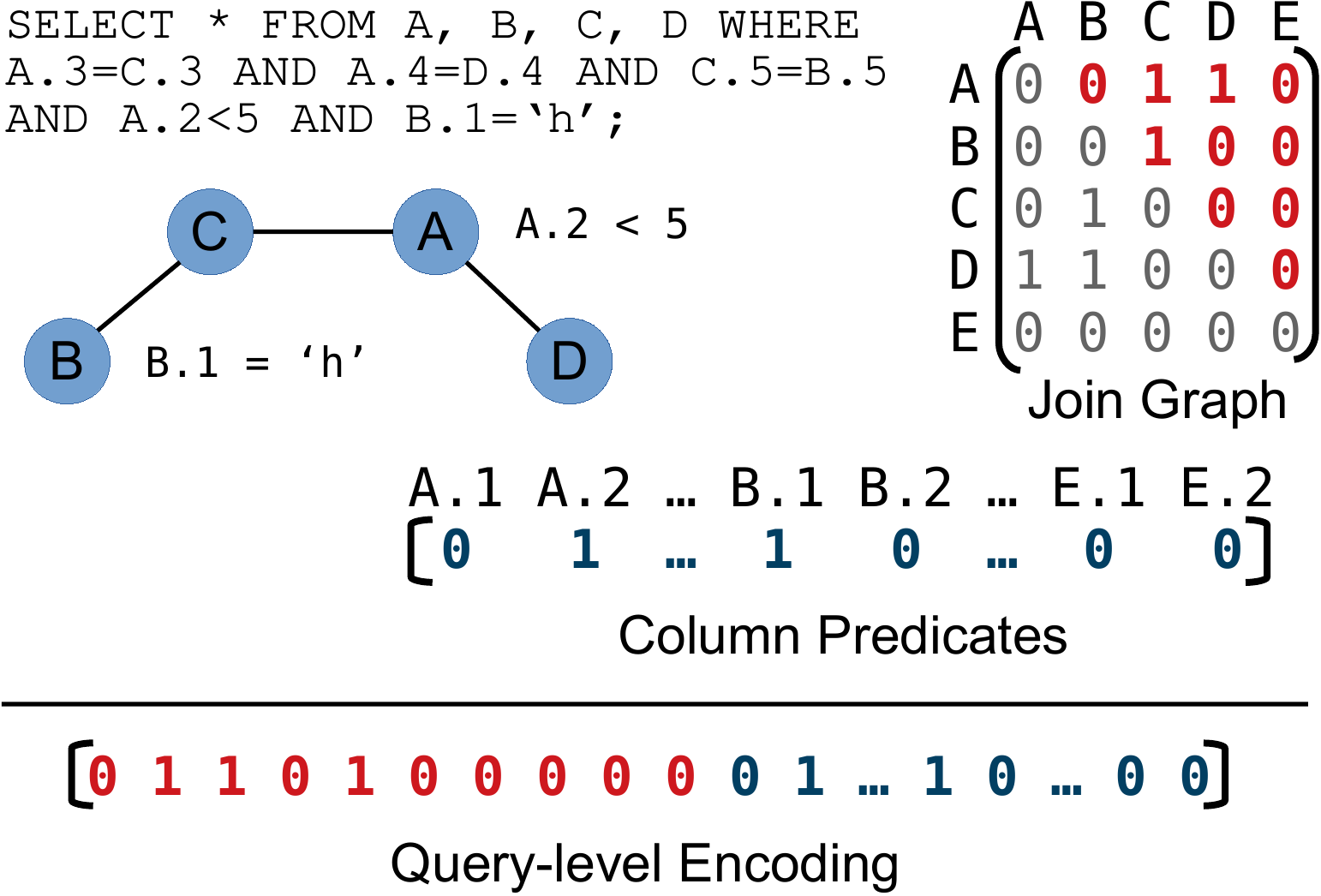}
    \caption{Query-level encoding}
    \label{fig:features}
\end{figure}

\label{sec:features}
\begin{figure}
    \centering
    \includegraphics[width=0.48\textwidth]{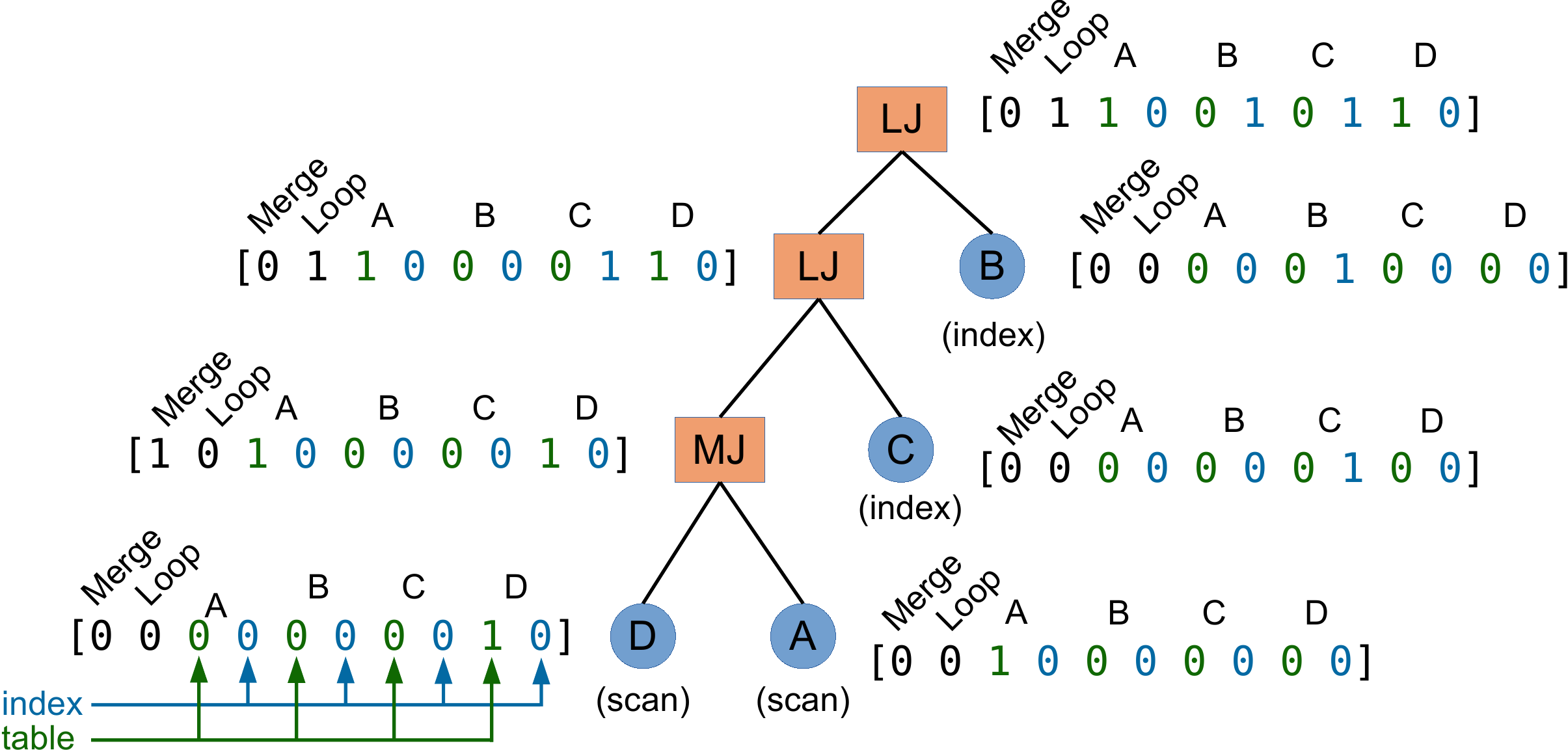}
    \caption{Plan-level encoding}
    \label{fig:tree_nomatrix}
\end{figure}

\begin{enumerate}
\item{\oh (the \emph{existence} of a predicate): is a simple  ``one-hot'' encoding of which attributes are involved in a query predicate. 
The length of the one-hot encoding vector is the number of attributes over all database tables.
For example, Figure~\ref{fig:features} shows the ``one-hot'' encoded vector with the positions for attribute $A.2$ and $B.1$ set to 1, since both attributes are used as part of predicate. Note that join predicates are not considered here.  
Thus, the learning agent \emph{only} knows whether an attribute is present in a predicate or not. While naive, the \oh representation can be built without any access to the underlying database.}
\item{\hist (the \emph{selectivity} of a predicate): is a simple extension of the previous one-hot encoding which replaces the indication of a predicate's existence with the predicted selectivity of that predicate (e.g., $A.2$ could be $0.2$, if we predict a selectivity of $20\%$).
For predicting selectivity, we use an off-the-shelf histogram approach with uniformity assumptions, as used by \PG and other open-source systems.}
\item{\rv (the \emph{semantics} of a predicate): is the most advanced encoding scheme, where we use \emph{row vectors}. We designed row vectors based on a natural language processing (NLP) model mirroring word vectors~\cite{word2vec}. In this case, each entry in the column predicate vector contains semantically relevant information related to the predicate. This encoding requires building a model over the data in the database, and is the most expensive option. We discuss row vectors in Section~\ref{sec:pari_features}.}
\end{enumerate}

The more powerful the encoding, the more degrees of freedom the model has to learn complex relationships. However, this does not necessarily mean that the model cannot learn more complex relationships with a simpler encoding. For example, even though \hist  does not encode anything about correlations between tables, the model might still learn about them and accordingly correct the cardinality estimations internally, e.g. from repeated observation of query latencies. 
However, with the \rv encoding, we make \system's job easier by providing a semantically-enhanced representation of the query predicate.  
% RM: did my pass, I like it.
%\tim{please double check. I am very certain, that it is true. But if there are any doubts, please let me know.}
%\nt{The last two sentences don't seem clear to me.}

\sparagraph{Plan Encoding} The second encoding we require is to represent the partial or complete query execution plan. 
While prior works~\cite{sanjay_wat, rejoin} have flattened the tree structure of each partial execution plan, our encoding \emph{preserves the inherent tree structure of execution plans}. We transform each node of the partial execution plan into a vector, creating a tree of vectors, as shown in Figure~\ref{fig:tree_nomatrix}. While the number of vectors (i.e., number of tree nodes) can increase, and the structure of the tree itself may change (e.g., left deep or bushy), every vector has the same number of columns.

These vectors are created as follows: each node is transformed into a vector of size $|J| + 2|R|$, where $|J|$ is the number of different join operators, and $|R|$ is the number of relations. The first $|J|$ entries of each vector encode the join type  (e.g., in Figure~\ref{fig:tree_nomatrix}, the root node uses a loop join), and the next $2|R|$ entries encode which relations are being used, and what type of scan (table, index, or unspecified) is being used. For leaf nodes, this subvector is a one-hot encoding, unless the leaf represents an unspecified scan, in which case it is treated as though it were both an index scan and a table scan (a 1 is placed in both the ``table'' and ``index'' columns). For any other node, these entries are the union of the corresponding children nodes. For example, the bottom-most loop join operator in Figure~\ref{fig:tree_nomatrix} has ones in the position corresponding to table scans over D and A, and an index scan over C.
  
Note that this representation can contain two partial query plans (i.e., several roots) which have yet to be joined, e.g. to represent the following partial plan:
\begin{equation*}
P = \left[(T(D) \bowtie_M T(A)) \bowtie_L I(C)\right], \quad \left[U(B)\right]
\end{equation*}

\noindent When encoded, the $U(B)$ root node would be encoded as:
\begin{equation*}
  [0 0 0 0 1 1 0 0 0 0]
\end{equation*}
Intuitively, partial execution plans are built ``bottom up'', and partial execution plans with multiple roots represent subplans that have yet to be joined together with a join operator. The purpose of these encodings is merely to provide a representation of execution plans to \system's value network, described next.

%% file: lo_value_network.tex
\section{Value Network}
\label{sec:value_network}

\begin{figure*}
  \centering
  \includegraphics[width=0.85\textwidth]{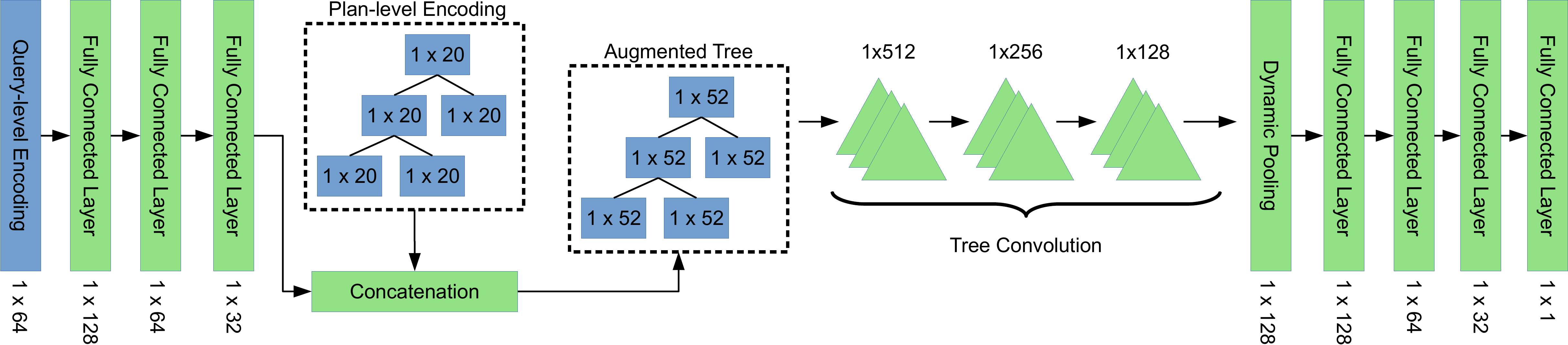}
  \vspace{-0.4cm}
  \caption{Value network architecture}
  \label{fig:network}
\end{figure*}

%\tim{Figure 3 value network should again use the same naming. Query Encoding, and Plan Encoding. Currently it talks about Vectorized Trees}

In this section, we present the \system \emph{value network}, a deep neural network model which is trained to approximate the {\em best-possible query latency} that a partial execution plan $P_i$ could produce (in other words, the best-possible query latency achievable by a complete execution plan $P_f$ such that $P_i$ is a subplan of $P_f$). 
%We note that this approach mirrors Q-learning~\cite{q}, for which we provide more mathematical intuition in Appendix~\ref{apx:q}. 
Since knowing the best-possible complete execution plan for a query ahead of time is impossible (if it were possible, the need for a query optimizer would be moot), we approximate the best-possible query latency with the best query latency \emph{seen so far by the system}.

Formally, let \system's \emph{experience} $E$ be a set of complete query execution plans $P_f \in E$ with known latency, denoted $L(P_f)$. We train a neural network model $M$ to approximate, for all $P_i$ that are a subplan of any $P_f \in E$:
\begin{equation*}
M(P_i) \approx \min \{ C(P_f) \mid P_i \subset P_f \land P_f \in E\}
\end{equation*}

%\nt{$C(P_f)$ below is the \emph{cost} of a "complete" plan, right? Also, shall we call it "learned cost", to avoid a potential mix-up with traditional cost functions and models.}

\noindent where $C(P_f)$ is the \emph{cost} of a complete plan. The user can change the cost function to alter the behavior of \system. For example, if the user is concerned only with minimizing total query latency across the workload, the cost could be defined as the latency, i.e.,
$$C(P_f) = L(P_f).$$
However, if instead the user prefers to ensure that every query $q$ in a workload performs better than a particular baseline, the cost function can be defined as
$$C(P_f) = L(P_f) / Base(P_f),$$
\noindent where $Base(P_f)$ is latency of plan $P_f$ with that baseline. Regardless of how the cost function is defined, \system will attempt to minimize it over time. We experimentally evaluate both of these cost functions in Section~\ref{sec:expr_opt}.

The model is trained by minimizing a loss function~\cite{dnn}. We use a simple L2 loss function:
\begin{equation*}
(M(P_i) - \min \{ C(P_f) \mid P_i \subset P_f \land P_f \in E\})^2.
\end{equation*}
%\noindent which penalizes the network for over or under predicting, and produces larger penalties for larger errors.

\sparagraph{Network Architecture} The architecture of the \system value network model is shown in Figure~\ref{fig:network}.\footnote{We omit activation functions, present between each layer, from our diagram and our discussion.} We designed the model's architecture to create an \emph{inductive bias}~\cite{inductive_bias_rl} suitable for query optimization: the structure of the neural network itself is designed to reflect an intuitive understanding of what causes query plans to be fast or slow. Intuitively, humans studying query plans learn to recognize suboptimal or good plans by pattern matching: a merge join on top of a hash join with a shared join key is likely inducing a redundant sort or hash step; a loop join on top of two hash joins is likely to be highly sensitive to cardinality estimation errors; a hash join using a fact table as the ``build'' relation is likely to incur spills; a series of merge joins that do not require re-sorting is likely to perform well, etc. Our insight is that all of these patterns can be recognized by analyzing subtrees of a query execution plan. Our model architecture is essentially a large bank of these patterns that are learned \emph{automatically, from the data itself}, by taking advantage of a technique called \emph{tree convolution}~\cite{tree_conv} (discussed in Section~\ref{sec:tree_conv}). %\hongzi{can move this intuition to tree conv section.} \ma{I'd probably keep it here but we don't really say how the architecture creates inductive bias? What specifically do we mean by this? Is it that tree structure matters, and tree convolution is a way of capturing parent-child relationships in the tree?}

As shown in Figure~\ref{fig:network}, when a partial query plan is evaluated by the model, the query-level encoding is fed through a number of fully-connected layers, each decreasing in size. The vector outputted by the third fully connected layer is concatenated with the plan-level encoding, i.e., each tree node (the same vector is added to all tree nodes). This is a standard technique~\cite{alphago} known as ``spatial replication''~\cite{bicyclegan} for combining data that has a fixed size (the query-level encoding) and data that is dynamically sized (the plan-level encoding). Once each tree node vector has been augmented, the forest of trees is sent through several tree convolution layers~\cite{tree_conv}, an operation that maps trees to trees. Afterwards, a dynamic pooling operation~\cite{tree_conv} is applied, flattening the tree structure into a single vector. Several additional fully connected layers are used to map this vector into a single value, used as the model's cost prediction for the inputted execution plan. A formal description of the value network model is given in Appendix~\ref{apx:model}.

\input{tree_conv}

%%% Local Variables:
%%% mode: latex
%%% TeX-master: "main"
%%% End:

%% file: tree_conv.tex
\subsection{Tree Convolution}
\label{sec:tree_conv}

% Tree convolution is like conv for images, but for trees. Each node has multiple channels of input and we get multiple channels of output.

% Sliding a filter over a tree to produce another tree

% Filters look at one parent and two children. Filters might represent ``am I joining table A with table B?'' (which might be a really bad or really good join) or ``are both of children sorted on the key that I need?'' (in which case merge join is a likely operator) or ``is this join likely to use very few rows from the right side'' (in which case you might want an index). stress: (1) this is an intuitive indcutive bias for the problem and (2) these features are learned automatically by the network via SGD.

% By applying multiple layers of tree conv, we learn more and more advanced features. 

\begin{figure*}
  \centering
  \includegraphics[width=0.9\textwidth]{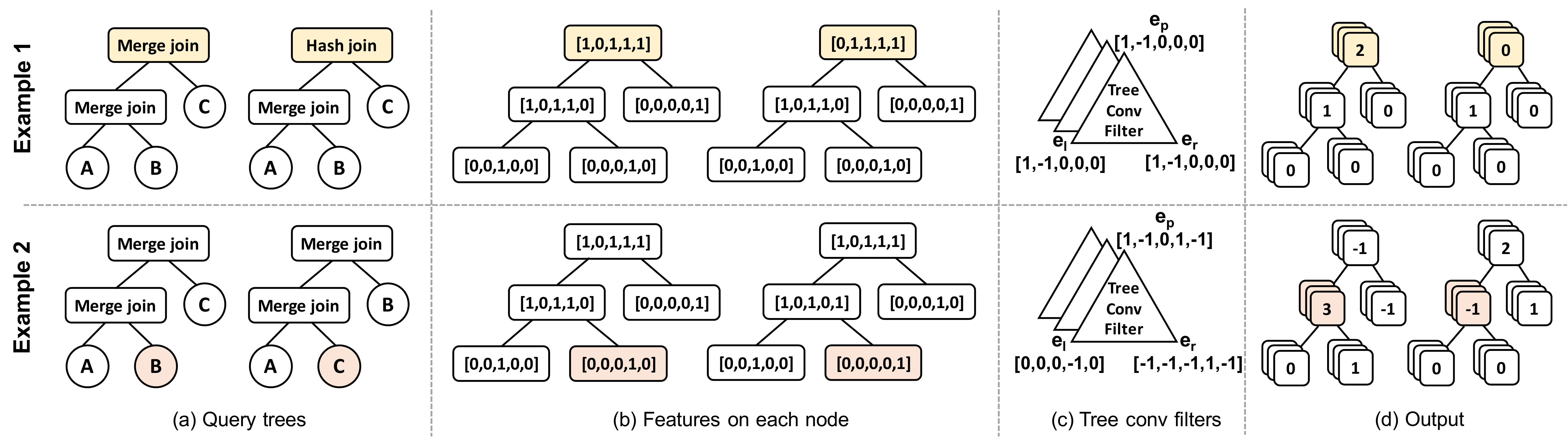}
  \vspace{-0.2cm}
  \caption{Tree convolution examples}
  \label{fig:tree_conv}
\end{figure*}

Common neural network models, like fully-connected neural networks or convolution neural networks, take as input tensors with a fixed structure, such as a vector or an image. In our problem, the features embedded in each execution plan are structured as nodes in a query plan tree (e.g., Figure~\ref{fig:tree_nomatrix}). To process these features, we use tree convolution methods~\cite{tree_conv}, an adaption of traditional image convolution for tree-structured data.

Tree convolution is a natural fit for this problem. Similar to the convolution transformation for images, tree convolution slides a set of \emph{shared} filters over each part of the query tree locally. Intuitively, these filters can capture a wide variety of local parent-children relations. For example, filters can look for hash joins on top of merge joins, or a join of two relations when a particular predicate is present. The output of these filters provides signals utilized by the final layers of the value network; filter outputs could signify relevant factors such as when the children of a join operator are sorted on the key (in which case merge join is likely a good choice), or a filter might estimate if the right-side relation of a join will have low cardinality (indicating that an index may be useful). We provide two concrete examples later in this section.

Operationally, since each node on the query tree has exactly two child nodes,\footnote{We attach nodes with all zeros to each leaf node.} each filter consists of three weight vectors, $e_p, e_l, e_r$. Each filter is applied to each local ``triangle'' formed by the vector $x_p$ of a node and two of its left and right child, $x_l$ and $x_r$ to produce a new tree node $x_p^\prime$:
\begin{equation*}
    x_p^\prime = \sigma(e_p \odot x_p + e_l \odot x_l + e_r \odot x_r).
\end{equation*}
Here, $\sigma(\cdot)$ is a non-linear transformation (e.g., ReLU~\cite{relu}), $\odot$ is a dot product, and $x_p'$ is the output of the filter. Each filter thus combines information from the local neighborhood of a tree node (its children).
%
% $f$ (e.g., matrix multiplication followed by a non-linear
% activation function) applied to the local “triangle” consisting of the vector featurization
% $x_p$ of a node and two of its children $x_l$ and $x_r$: $x_p' = f(x_p, x_l, x_r)$. 
% %
% Concretely, we dot product the up-left-right concatenation of the weight vectors in the
% tree convolution filter with the parent-left-right concatenation of feature vectors for each node on the query tree, followed by a ReLU non-linear transformation~\cite{relu}.
% %
% The output of the transformation $x_p'$ captures the local summarization of an
% intermediate join. 
%
The same filter is ``slid'' across each tree in a execution plan, allowing a filter to be applied to execution plans with arbitrarily sized trees. A set of filters can be applied to a tree in order to produce another tree with the same structure, but with potentially different sized vectors representing each node.
In a large neural network, such as those in our experimental evaluation, typically hundreds of filters are applied.
%
%\ma{giving some idea of how many such filters we have at each node would be helpful.}

Since the output of a tree convolution is also a tree with the same shape as the input (but with different sized vector representing each node), multiple layers of tree convolution filters can be sequentially applied to an execution plan. The first layer of tree convolution filters will access the augmented execution plan tree, and each filter will ``see'' each parent/left child/right child triangle of the original tree. The amount of information seen by a particular filter is called the filter's \emph{receptive field}~\cite{cnn_seg}. The second layer of convolution filters will be applied to the output of the first, and thus each filter in this second layer will see information derived from a node $n$ in the original augmented tree, $n$'s children, and $n$'s grandchildren. Thus, each tree convolution layer has a larger receptive field than the last. As a result, the first tree convolution layer will learn simple features (e.g., recognizing a merge join on top of a merge join), whereas the last tree convolution layer will learn complex features (e.g., recognizing a left-deep chain of merge joins).

We present two concrete examples in Figure~\ref{fig:tree_conv} that show how the first layer of tree convolution can detect interesting patterns in query execution plans.
In Example 1 of Figure~\ref{fig:tree_conv}a, we show two execution plans that differ only in the topmost join operator (a merge join and hash join).
As depicted in the the top portion of Figure~\ref{fig:tree_conv}b, the join type (hash or merge) is encoded in the first two bits of the feature vector in each node.
%
% Next, we dot product the up-left-right concatenation of the weight vectors in the
% tree convolution filter with the parent-left-right concatenation of feature vectors for each node on the query tree, followed by a ReLU non-linear transformation~\cite{relu}. \ma{Things like this would be much easier to understand in equations.}
%
Now, if a tree convolution filter (Figure~\ref{fig:tree_conv}c top) is
comprised of three weight vectors with $\{1, -1\}$ in the first two positions and zeros for the rest, it will serve as a ``detector'' for query plans with two merge joins in a row. This can be seen in Figure~\ref{fig:tree_conv}d (top): the root node of the plan with two merge joins in a row receives an output of 2 from this filter, whereas the root node of the plan with a hash join on top of a merge join receives an output of 0. Subsequent tree convolution layers can use this information to form more complex predicates, e.g. to detect three merge joins in a row (a pipelined query execution plan), or a mixture of merge joins and hash joins (which may require frequent re-hashing or re-sorting). 
In Example 2 of Figure~\ref{fig:tree_conv}, suppose A and B are physically sorted on the same key, and are thus optimally joined together with a merge join operator, but that C is not physically sorted. The tree convolution filter shown in Figure~\ref{fig:tree_conv}(c, bottom) serves as a detector for query plans that join A and B with a merge join, behavior that is likely desirable. The top weights recognize the merge join (\{1, -1\} for the first two entries) and the right weights prefer table B over all
other tables. The result of this convolution (Figure~\ref{fig:tree_conv}d bottom) shows its highest output for the merge join of A and B (in the first plan), and a negative output for the merge join of A and C (in the second plan).

In practice, filter weights are \emph{learned} over time in an end-to-end fashion. By using tree convolution layers in a neural network, performing gradient descent on the weights of each filter will cause filters that correlate with latency (e.g., helpful features) to be rewarded (remain stable), and filters with no clear relationship to latency to be penalized (pushed towards more useful values). This creates a corrective feedback loop, resulting in the development of filterbanks that generate useful features~\cite{deep_survey}.

%\tim{if we need space, we can probably remove the example. I like it, but I am not sure how many of our reviewers will find it helpful as tree convolution is not easy to grasp. }
% RM: I think I disagree. I don't think we hand-wave tree convolution away... we need to convince reviewers that it is a good inductive bias for the problem. If they can't grasp that, our only hope is that they'll be impressed with the results (which isn't a bad bet, but let's cover our bases).
% \ma{I like the examples. I'd suggest keeping this.}
%

%
% Notice that we do not hard-code the functionality of the filters---the weights on the filters are trained automatically end-to-end.

%%% Local Variables:
%%% mode: latex
%%% TeX-master: "main"
%%% End:

%% file: lo_search.tex
\subsection{DNN-Guided Plan Search}
\label{sec:search}

The value network predicts the quality (cost) of an execution plan, but it does not directly give an execution plan. Following several recent works in reinforcement learning~\cite{alphago, rl_search}, we combine the value network with a search technique to generate query execution plans, resulting in a \emph{value iteration technique~\cite{bellman}} (discussed at the end of the section).

Given a trained value network and an incoming query $q$, \system performs a search of the plan space for a given query. Intuitively, this search mirrors the search process used by traditional database optimizers, with the trained value network taking on the role of the database cost model. Unlike these traditional systems, the value network does not predict the cost of a subplan, but rather the best possible latency achievable from an execution plan that includes a given subplan. This difference allows us to perform a best-first search~\cite{best_first} to find an execution plan with low expected cost. Essentially, this amounts to repeatedly exploring the candidate with the best predicated cost until a halting condition occurs.

The search process for query $q$ starts by initializing an empty min heap to store partial execution plans. This min heap is ordered by the value network's estimation of a partial plan's cost. Then, a partial execution plan with an unspecified scan for each relation in $R(q)$ is added to the heap. For example, if $R(q) = \{A, B, C, D\}$, then the heap is initialized with $P_0$:
\begin{equation*}
P_0 = [U(A)], \quad [U(B)], \quad [U(C)], \quad [U(D)],
\end{equation*}
where $U(r)$ is the unspecified scan for the relation $r \in R(q)$.

At each search iteration, the subplan $P_i$ at the top of the min heap is removed. We enumerate all of $P_i$'s children, $Children(P_i)$, scoring them using the value network and adding them to the min heap. Intuitively, the children of $P_i$ are all the plans creatable by specifying a scan in $P_i$ or by joining two trees of $P_i$ with a join operator. Formally, we define $Children(P_i)$ as the empty set if $P_i$ is a complete plan, and otherwise as all possible subplans $P_j$ such that $P_i \subset P_j$ and such that $P_j$ and $P_i$ differ by either (1) changing an unspecified scan to a table or index scan, or (2) merging two trees using a join operator. Once each child is scored and added to the min heap, another search iteration begins, resulting in the next most promising node being removed from the heap and explored.

While one could terminate this search process as soon as a leaf node (a complete execution plan) is found, this search procedure can easily be transformed into an anytime search algorithm~\cite{anytime}, i.e. an algorithm that continues to find better and better results until a fixed time cutoff. In this variant, the search process continues exploring the most promising nodes from the heap until a time threshold is reached, at which point the most promising complete execution plan is returned. This gives the user control over the tradeoff between planning time and execution time. Users could even select a different time cutoff for different queries depending on their needs. In the event that the time threshold is reached before a complete execution plan is found, \system's search procedure enters a ``hurry up'' mode~\cite{vertica_opt}, and greedily explores the most promising children of the last node explored until a leaf node is reached. The cutoff time should be tuned on a per-application bases, but we find that a value of 250ms is sufficient for a wide variety of workloads (see Section~\ref{sec:expr_search}), a value that is acceptable for many applications.

From a reinforcement learning point of view, the combination of the value network with a search procedure is a \emph{value iteration} technique~\cite{bellman}. Value iteration techniques cycle between two steps: estimating the value function, and then using that value function to improve the policy. We estimate the value function via supervised training of a neural network, and we use that value function to improve a policy via a search technique. Q learning~\cite{q} and its deep neural network variants~\cite{dqn}, which have been recently used for query optimization~\cite{sanjay_wat}, are also value iteration methods: the value estimation step is similar, but they use that value function to select actions greedily (i.e., without a search). This approach is equivalent to \system's ``hurry up'' mode. As our experiments show, combining value estimation  with a search procedure leads to a system that is less sensitive to noise or inaccuracies in the value estimation model, resulting in significantly better query plans. This improvement has been observed in other fields as well~\cite{alphago, rl_search}.

%\tim{I think early on in this section, we should point out how the technique relates to the Berkeley paper and what makes it different}

%\ma{I'd actually briefly mention Q learning and other value-based methods when discussing the search procedure. I think broadly our scheme can be thought of as a value iteration algorithm. Value iteration methods conceptually iterate between estimating a value function, using it to improve the policy, again estimating the value function, and so on. These methods differ in how they derive an improved policy from the value function of a previous one. Q-learning picks actions greedily based on the value function. This can be sensitive to noise/inaccuracies in the value function model. Search on top of the value function is a stronger policy improvement operator (e.g., see explanation in AlphaGo).}

%%% Local Variables:
%%% mode: latex
%%% TeX-master: "main"
%%% End:

%% file: pari_features.tex
\section{Row Vector Embeddings}
\label{sec:pari_features}

\system can represent query predicates in a number of ways, including a simple one-hot encoding (\oh) or a histogram-based representation (\hist), as described in Section~\ref{sec:features}. Here, we motivate and describe \emph{row vectors}, \system's most advanced option for representing query predicates (\rv). 

Cardinality estimation is one of the most important problems in query optimization today~\cite{qo_unsolved, howgood}. Estimating cardinalities is directly related to estimating selectivities of query predicates -- whether these predicates involve a single table or joins across multiple tables. The more columns or joins are involved, the harder the problem becomes. 
%It is challenging to precisely capture semantic relationships among columns, which typically manifest themselves in the form of statistical correlations. 
Modern database systems make several simplifying assumptions about these correlations, such as uniformity, independence, and/or the principle of inclusion~\cite{job2}. These assumptions often do not hold in real-world workloads, causing orders of magnitude increases in observed query latencies~\cite{howgood}. In \system, we take a different approach: instead of making simplifying assumptions about data distributions and attempting to directly estimate predicate cardinality, we build a semantically-rich, vectorized representation of query predicates that can serve as an input to \system's value model, enabling the network to learn generalizable data correlations.

While \system supports several different encodings for query predicates, here we present \emph{row vectors}, a new technique based on the popular word2vec~\cite{word2vec} algorithm. Intuitively, we build a vectorized representation of each query predicate \emph{based on data in the database itself}. These vectors are meaningless on their own, but the \emph{distances between these vectors will have semantic meaning}. \system's value network can take these row vectors are inputs, and use them to identify correlations within the data and predicates with syntactically-distinct but semantically-similar values (e.g. $Genre$ is  ``action'' and $Genre$ is ``adventure'').

\subsection{{R-Vector} Featurization}

 The basic idea behind our approach is to capture contextual cues among values that appear in a database. To give a high-level example from the IMDB movie dataset~\cite{howgood}, if a keyword ``marvel-comics'' shows up in a query predicate, then we wish to be able to predict what else in the database would be relevant for this query (e.g., other Marvel movies).

\sparagraph{Word vectors} To generate row vectors, we use word2vec ---\,a natural language processing technique for embedding contextual information about collections of words~\cite{word2vec}. In the word2vec model, each sentence in a large body of text is represented as a collection of words that share a context, where similar words often appear in similar contexts. These words are mapped to a vector space, where the angle and distance between vectors reflect the similarity between words. For example, the words ``king'' and ``queen'' will have similar vector representations, as they frequently appear in similar contexts (e.g. ``Long live the...''), whereas words like ``oligarch'' and ``headphones'' will have dissimilar vector representations, as they are unlikely to appear in similar contexts. 

We see a natural parallel between sentences in a document and rows in a database: values of correlated columns tend to appear together in a database row. In fact, word2vec-based embeddings have recently been applied to other database problems, such as semantic querying~\cite{semantic-queries}, entity matching~\cite{deep_entity}, and data discovery~\cite{termite}. In \system, we use an off-the-shelf word2vec implementation~\cite{gensim} to build an embedding of each value in the database. We then utilize these embeddings to encode correlations across columns.\footnote{Predicates with comparison operators, e.g. {\tt IN} and {\tt LIKE}, can lead to multiple matches. In this case, we take the mean of all the matched word vectors as the embedding input.}

We explored two variants of this featurization scheme. In the first approach, we treat each row in every table of the database as a training sentence. This approach captures correlations within a table. To better capture cross-table correlations, in our second approach, we augment the set of training sentences by {\em partially denormalizing} the database. Concretely, we join large fact tables with smaller tables which share a foreign key, and each resulting row becomes a training sentence.
% This way, we can learn a compact representation for column values (e.g., of less than $20\%$ of the size of IMDB database) by training a word2vec model within a few hours. 
Denormalization enables learning correlations such as ``Actors born in Paris are more likely to play in French movies". While a Paris-born actor's name and birthplace may be stored only in the {\tt names} table, the {\tt cast\_info} table captures information relating this actor to many French movies. By joining these two tables together, we can make these relationships\,---\,such as the actor's name, birthplace, and all the French movies they have appeared in\,---\,explicit to the word2vec training algorithm.
Conceptually, denormalizing the entire database would provide the best embedding quality, but at the expense of significantly increasing the size and number of training examples (a completely denormalized database may have trillions of rows), and hence the word2vec training time. A fully denormalized database is often unfeasible to materialize. In Section~\ref{sec:training_time}, we present the details of the row vector training performance in practice. 
Our word2vec training process is open source, and available on GitHub.\footnote{\small \url{https://github.com/parimarjan/db-embedding-tools}}

\begin{figure}
  \centering
  \begin{subfigure}{0.235\textwidth}
    \includegraphics[width=\textwidth]{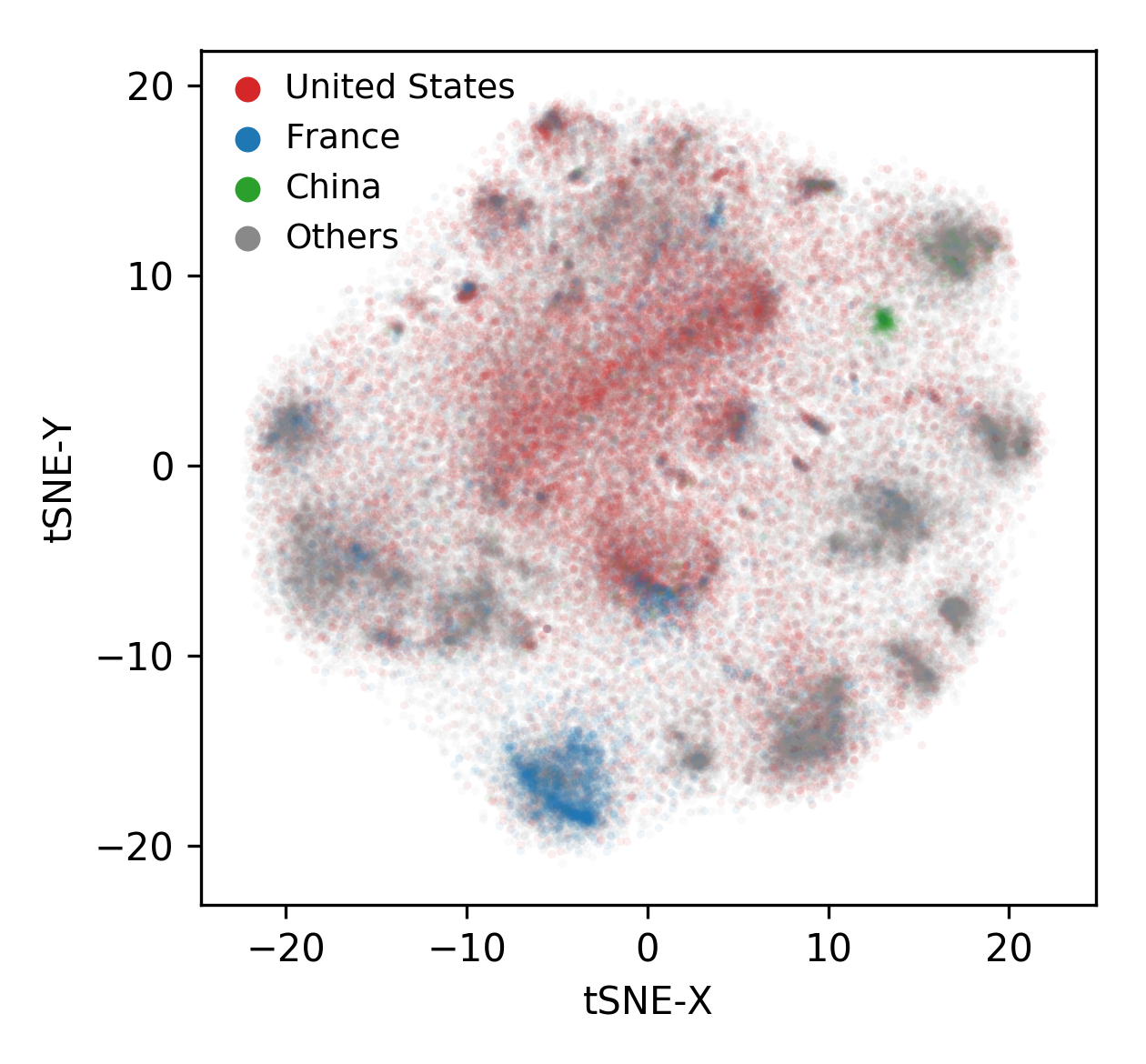}
    \caption{
    Birthplace of each actor
    }
    \label{fig:actor_country}
  \end{subfigure}
  \begin{subfigure}{0.235\textwidth}
    \includegraphics[width=\textwidth]{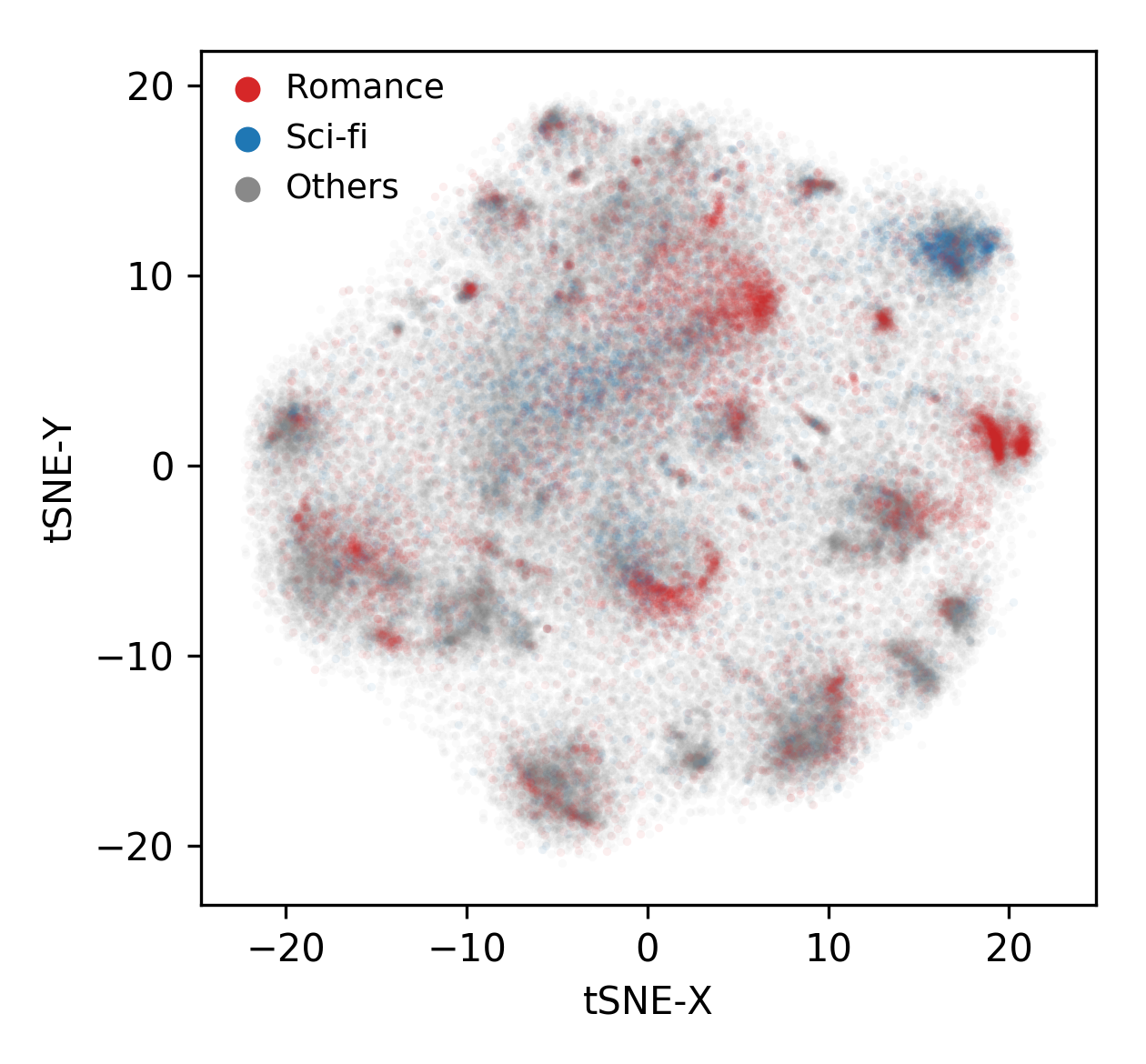}
    \caption{
    Top actors in each genre
    % Shows the top $10,000$ actors in these genres, ranked by the number of movies in that genre the actor has been part of. This relationships spans five tables in the IMDB database: name, title, movie\_info, info\_type.
    }
    \label{fig:actor_genre}
  \end{subfigure}
\caption{t-SNE projection of actor names embedded in the word2vec model. Column correlations across multiple IMDB tables show up as semantically meaningful clusters.}
\label{fig:tsne}
\end{figure}

Figure~\ref{fig:tsne} presents a visual example to show how row vectors capture semantic correlations between database tables. We use t-SNE~\cite{tsne} to project embedded vectors of actor names from their original 100-dimensional space into two-dimensional space for plotting. The t-SNE algorithm finds low dimensional embeddings of high dimensional spaces that attempts to maintain the distance between points: points that are close together in the two-dimensional space are close together in the high dimensional space as well, and points that are far apart in the low dimensional space are far apart in the high dimensional space as well. 

As shown, various semantic groups (e.g., Chinese actors, sci-fi movie actors) are clustered together (and are thus also close together in the original high-dimensional space), even when these semantic relationships  span multiple tables.
Intuitively, this provides helpful signals to estimate query latency given similar predicates: as many of the clusters in Figure~\ref{fig:tsne} are linearly separable, their boundaries can be  learned by machine learning algorithms. In other words, since predicates with similar \emph{semantic} values (e.g., two American actors) are likely to have similar \emph{correlations} (e.g., be in American films), representing the semantic value of a query predicate allows the value network to recognize similar predicates as similar.
In Section~\ref{sec:eval-feature}, we find that row vectors consistently improves \system's performance compared to other featurizations.

\sparagraph{Row vector construction} In our implementation, the feature vectors for query predicates are constructed as follows. For every distinct value in the underlying database, we generate vectors which are a concatenation of the following:
\begin{enumerate}
\item One-hot encoding of the comparison operators (e.g. equal or not equal to)
\item Number of matched words
\item Column embedding generated by word2vec (100 values, in our experiments)
\item Number of times the given value is seen in the training
\end{enumerate}

The concatenated vectors replace the ``1''s or ``0''s in the column predicate vector of \oh representation of the query-level information (see Section~\ref{sec:lo}). For columns without a predicate, zeros are added so that the vector remains the same size regardless of the number of predicates.
%We experimented with different variations of 1-4 above; we report on our findings in Section~\ref{sec:eval-feature}.

% Please note that predicates with comparison operators {\tt IN} and {\tt LIKE} can lead to multiple matches. In such cases, we take the mean of all the matched word vectors as the embedding input. If the predicate was {\tt LIKE \%A\%}, then this clearly does not represent anything useful. But if there were only a few similar words that matched the predicate, e.g., {\tt LIKE \%spider\%}, then this seems more sensible. In case there was no match in the trained model, then we just output a vector of 0s. This could happen if the word was really rare in the database, or it was not in the database at all.

%\ma{Have we tried to see what happens if you keep 1, 2, and 4, but don't include 3 (the word2vec embedding)? Does that hurt performance?}
%\nt{Should the outcome of the variations you tried be briefly summarized here?}

\subsection{Analysis}

Here, we analyze the embedding space learned by our row vector approach on the IMDB dataset.
% It ranges from -1 to +1, where values between 0 and 1 represent increasing levels of positive correlation, 0 represents uncorrelated values, and values between -1 to 0 represent negative correlations. 
We use the below SQL query from the IMDB database to illustrate how learned row vectors can be useful for tasks like cardinality estimation and query optimization.

% \begin{figure}
% \scalebox{1.0}
% {
% \begin{lstlisting}[
%           language=SQL,
%           showspaces=false,
%           basicstyle=\ttfamily,
%           numbers=none,
%           numberstyle=\tiny,
%           commentstyle=\color{gray}
%         ]
%   SELECT count(*)
%   FROM title as t,
%       movie_keyword as mk,
%       keyword as k,
%       info_type as it,
%       movie_info as mi
%   WHERE it.id = 3
%   AND it.id = mi.info_type_id
%   AND mi.movie_id = t.id
%   AND mk.keyword_id = k.id
%   AND mk.movie_id = t.id
%   AND k.keyword ILIKE '%love%'
%   AND mi.info ILIKE '%romance%'
% \end{lstlisting}
% \caption{Example query with correlations}
% \label{fig:sql}
% }
% \end{figure}

This query counts the number of movies with genre ``romance'' and containing the keyword ``love''. It spans five tables in the IMDB dataset. As input to word2vec training, we partially denormalized these tables by joining {\tt title, keyword\_info, keyword} and {\tt title, movie\_info, info\_type}. It is important to note that, after this denormalization, keywords and genres do not appear in the same row, but keywords-titles as well as titles-genres do appear in separate rows.

\begin{lrbox}{\mybox}%
    \begin{lstlisting}%
    [
           language=SQL,
           showspaces=false,
           basicstyle=\ttfamily,
           numbers=none,
           numberstyle=\tiny,
           commentstyle=\color{gray}
        ]
  SELECT count(*)
  FROM title as t,
       movie_keyword as mk,
       keyword as k,
       info_type as it,
       movie_info as mi
  WHERE it.id = 3
  AND it.id = mi.info_type_id
  AND mi.movie_id = t.id
  AND mk.keyword_id = k.id
  AND mk.movie_id = t.id
  AND k.keyword ILIKE '%love%'
  AND mi.info ILIKE '%romance%'
  \end{lstlisting}%

\end{lrbox}%

\begin{figure}
\hspace{1.3cm}
\scalebox{0.8}{\usebox{\mybox}}
\caption{Example query with correlations}
\label{fig:sql}
\end{figure}

\begin{table}
\centering
\begin{tabular}{|c|c|c|c|}
\hline
    %   &       & Cosine     & True \\
Keyword & Genre & Similarity & Cardinality \\
\hline
love & romance & 0.24 & 11128 \\
love & action & 0.16 & 2157 \\
love & horror & 0.09 & 1542 \\
% screaming & horror & 0.23 & 376 \\
% screaming & romance & 0.13 & 87 \\
% screaming & action & 0.11 & 79 \\
fight & action & 0.28 & 12177 \\
fight & romance & 0.21 & 3592 \\
fight & horror & 0.05 & 1104 \\
% joke & comedy & 0.23 & 1833 \\
% joke & romance & 0.23 & 112 \\
\hline
\end{tabular}
\caption{Similarity vs. Cardinality. In this case, correlated keywords and genres, as shown in the SQL query in Figure~\ref{fig:sql}, also have higher similarity and higher cardinality.}
%\caption{Here we modify the keyword and genre in the SQL query shown in \ref{keyword-genre-sql}. Postgres, with its uniformity assumptions, would always estimate the cardinality to be close to $1$. Meanwhile, the real cardinalities vary wildly. For some intuitive examples, we see that the cosine distance in the embeddings seem to capture the correlation effects. The last couple of rows show that this is not very precise: despite having the same cosine distance, the cardinalities are still an order of magnitude apart.}
\label{tab:example}
\end{table}

In Table~\ref{tab:example}, we compare the cosine similarity (higher value indicating higher similarity) between the vectors for keywords and genres to their true cardinalities in the dataset. As shown, highly correlated keywords and genres (e.g., ``love'' and ``romance'') have higher cardinalities. As a result, this embedding provides a representative feature that can somewhat substitute a precise cardinality estimation: a model built using these vectors as input can learn to understand the correlations within the underlying table.
% Also, similarity values follow a similar trend as cardinalities. 
% In fact, they only disagree in one pair of instances: joke-comedy vs. joke-romance.
% We believe that with a complete denormalization of all 5 tables, we could have captured stronger semantics between keywords and genres, but 
% These numbers indicate that our embeddings capture transitive relationships.
%

%It is also not necessary for such a learning agent to actually see the ``love'' and ``romance'' query in its training set - perhaps instead, it saw a query with ``fight'' and ``action'' in the training. The cosine distance between the embedded vectors for ``fight'' and ``action'' is 0.28, thus seeing that those predicates were correlated would give the learning agent useful information about completely unseen predicates such as ``romance'' and ``love''.
%\nt{I didn't quite understand the argument above.}

PostgreSQL, with its uniformity and independence assumptions, always estimates the cardinalities for the final joined result to be close to 1, and therefore prefers to use nested loop joins for this query. In reality, the real cardinalities vary wildly, as shown in Table \ref{tab:example}.
%This was one of the queries in the Ext-Job dataset (set of extra, unseen queries on the JOB dataset), and
For this query, \system decided to use hash joins instead of nested loop joins, and as a result, was able to execute this query $60\%$ faster than PostgreSQL. %by using hash joins instead of nested loop joins.

This simple example provides a clear indication that row vector embeddings can capture meaningful relationships in the data beyond histograms and one-hot encodings. This, in turn, provides \system with useful information in the presence of highly correlated columns and values. An additional advantage of our row embedding approach is that, by learning semantic relationships in a given database, \system can gain useful information even about column predicates that it has never seen before in its training set (e.g., infer similar cardinality using similar correlation between two attributes).

While we can observe useful correlations in the row vectors built for the IMDB dataset, language models like word2vec are notoriously difficult to interpret~\cite{wordvec3}. To the best of our knowledge, there are no formal methods to ensure that a word2vec model -- on either natural language or database rows -- will always produce helpful features. Developing such formal analysis is an active area of research in machine learning~\cite{wordvec1,wordvec2}. Thus, while we have no evidence that our row vector embedding technique will work on every imaginable database, we argue that our analysis on the IMDB database (a database with significant correlations / violations of uniformity assumptions) provides early evidence that row vectors may also be useful in other similar applications with semantically rich datasets. We plan to pursue this as part of our future work.
%At worst, the row vector encoding will perform no worse than the one-hot encoding, as row vectors are ensured to be non-zero~\cite{word2vec}. 

%%% Local Variables:
%%% mode: latex
%%% TeX-master: "main"
%%% End:

%% file: experiments.tex
\section{Experiments}
\label{sec:experiments}

We evaluated \system's performance using both synthetic and real-world datasets to answer the following questions: (1) how does the performance of \system compare to commercial, high-quality optimizers, (2) how well does the optimizer generalize to new queries, (3) how long is the optimizer execution and training time, (4) how do the different encoding strategies impact the prediction quality, (5) how do other parameters (e.g., search time or loss function) impact the overall performance, and finally, (6) how robust is \system to estimation errors.

\subsection{Setup}

We evaluate \system across a number of different database systems, using three different benchmarks:
\begin{enumerate}
\item{\JOB: the join order benchmark~\cite{howgood}, with a set of queries over the Internet Movie Data Base (IMDB) consisting of complex predicates, designed to test query optimizers.}
\item{\TPC: the standard TPC-H benchmark~\cite{tpch}, using a scale factor of 10.}
\item{\Vertica: a 2TB dataset together with 8,000 unique queries from an internal dashboard application, provided by a large corporation (on the condition of anonymity).}
\end{enumerate}

Unless otherwise stated, all experiments are conducted by randomly placing 80\% of the available queries into a training set, and using the other 20\% of the available queries as a testing set. In the case of TPC-H, we generated 80 training and 20 test queries based on the benchmark query templates without reusing templates between training and test queries. 

We present results as the median performance from fifty randomly initialized neural networks. The Adam~\cite{adam} optimizer is used for network training, as well as layer normalization~\cite{layer_norm} to stabilize neural network training. The ``leaky'' variant of rectified linear units~\cite{relu} are used as activation functions. We use a search time cutoff of 250ms. The network architecture follows Figure~\ref{fig:network}, except the size of the plan-level encoding is dependent on the featurization chosen (e.g. \oh or \hist).
%We decompose our experimental evaluation into several parts. In Section~\ref{sec:expr_pg}, we evaluate \system's ability to improve upon the \PG optimizer. Section~\ref{sec:expr_other_engines} explores \system's behavior when bootstrapped using the \PG optimizer, but then applied to other engines. In Section~\ref{sec:expr_perf}, we summarize and analyze the final performance characteristics of \system across each DBMS and each featurization. Section~\ref{sec:only_joins} shows how \system's behavior changes when \system is only responsible for producing a join ordering, and the underlying DBMS handles operator and index selection. In Section~\ref{sec:expr_opt}, we analyze the effects of \system's optimization metric. Section~\ref{sec:expr_search} evaluates our search technique. Finally, Section~\ref{sec:overhead} analyzes \system's runtime overhead.

\subsection{Overall Performance}

To evaluate \system's overall performance, we compared the mean execution time of the query plans generated by \system on two open-source (PostgreSQL 11, SQLite 3.27.1), and two commercial (Oracle 12c, Microsoft SQL Server 2017 for Linux) database systems, with the execution time of the plans generated by each system's native optimizer, for each of our three workloads. Due to the license terms~\cite{dewitt_clause} of Microsoft SQL Server and \Ora, we can only show performance in relative terms.

For initial experience collection for \system, we always used the \PG optimizer as the expert. That is, for every query in the training set, we used the \PG optimizer to generate an initial query plan.  We then measured the execution time of this plan on the target execution engine (e.g., \MS) by forcing the target system, through query hints, to obey the proposed query plan.  Next, we directly begin training: \system encodes the execution plan for each query in the training set, these plans are executed on the native system, and the encoded vectors along with the resulting run times are added to \system's experience.

\begin{figure}
  \centering
  \includegraphics[width=0.44\textwidth]{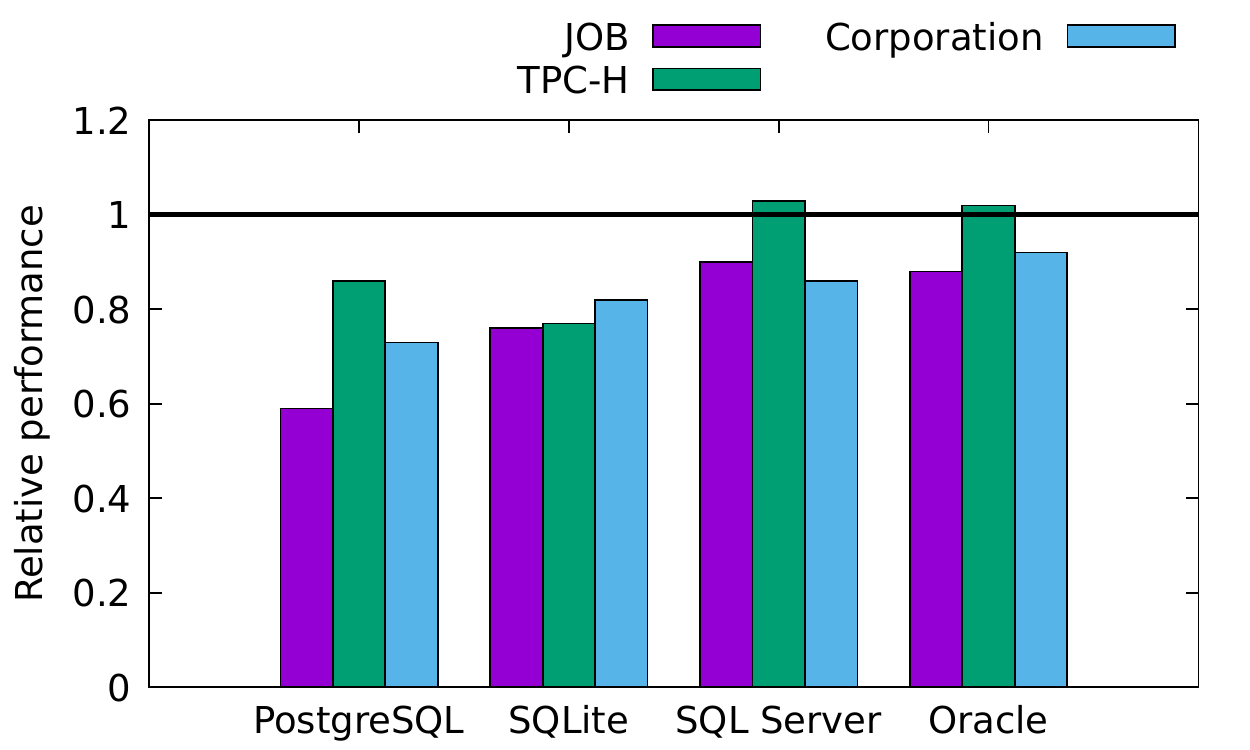}
  \caption{Relative query performance to plans created by the native optimizer (lower is better) for different workloads}
  \label{fig:trained}
\end{figure}

Figure~\ref{fig:trained} shows the relative performance of \system after 100 training iterations on each test workload, using the \rv encoding over the holdout dataset (lower is better). 
For example, with \PG and the \JOB workload, \system produces queries that take only 60\% of average execution time than the ones created by the original \PG optimizer. 
Since the \PG optimizer is used to gather initial expertise for \system, this demonstrates \system's ability to improve upon an existing open-source optimizer.

Moreover, for \MS and the \JOB and \Vertica workloads, the query plans produced by \system are also 10\% faster than the plans created by the commercial optimizers on their native platforms. 
Importantly, both commercial optimizers, which include a multi-phase search procedure and a dynamically-tuned cost model with hundreds of inputs ~\cite{inside_mssql, testing_mssql}, are expected to be substantially more advanced than \PG's optimizer. Yet, by bootstrapping only with \PG's optimizer, \system is able to eventually outperform or match the performance of these commercial optimizers on their own platforms.
% This is impressive, as it is expected that both commercial optimizers are substantially more advanced than \PG's optimizer, including a multi-phase search procedure and a dynamically-tuned cost model with hundreds of inputs ~\cite{inside_mssql, testing_mssql}.
Note that the faster execution times are solely based on better query plans without run-time modifications of the system.
% Furthermore, the boostrapping is only done with \PG's optimizer, so \system did not have access to the query plans generated by the other optimizers for training, yet still outperforms them on their own platforms.
The only exception where \system does not outperform the two commercial systems is for the TPC-H workload. We suspect that both \MS and \Ora were overtuned towards TPC-H, as it is one of the most common benchmarks.

Overall, this experiment demonstrates that {\bf \system is able to create plans, which are as good as, and sometimes even better than, open-source optimizers and their significantly superior commercial counterparts.} However, Figure~\ref{fig:trained} only compares the median performance of \system after the 100th training episode.
This naturally raises the following questions: (1) how does the performance compare with a fewer number of training episodes and how long does it take to train the model to a sufficient quality (answered in the next subsection), and (2) how robust is the optimizer to various imputations (answered in Section~\ref{sec:exp:robustness}). 

\subsection{Training Time}
\label{sec:exp:training_time}

To analyze the convergence time, we measured the performance after every training iteration, for a total of 100 complete iterations.
We first report the learning curves in training intervals to make the different systems comparable (e.g., a training episode with \MS might run much faster than \PG).
Afterwards, we report the wall-clock time to train the models on the different systems.
Finally, we answer the question of how much our bootstrapping method helped with the training time. 

\subsubsection{Learning Curves}
\label{sec:expr_pg}

\begin{figure*}[t]
  \begin{tabular}{m{1mm} m{0.22\textwidth} m{0.22\textwidth} m{0.22\textwidth} m{0.22\textwidth}}
    & \centering \PG & \centering \SQLite
    & \centering \MS & \begin{center}\Ora\end{center} \\
    \rotatebox{90}{\JOB}
    & \includegraphics[width=0.22\textwidth]{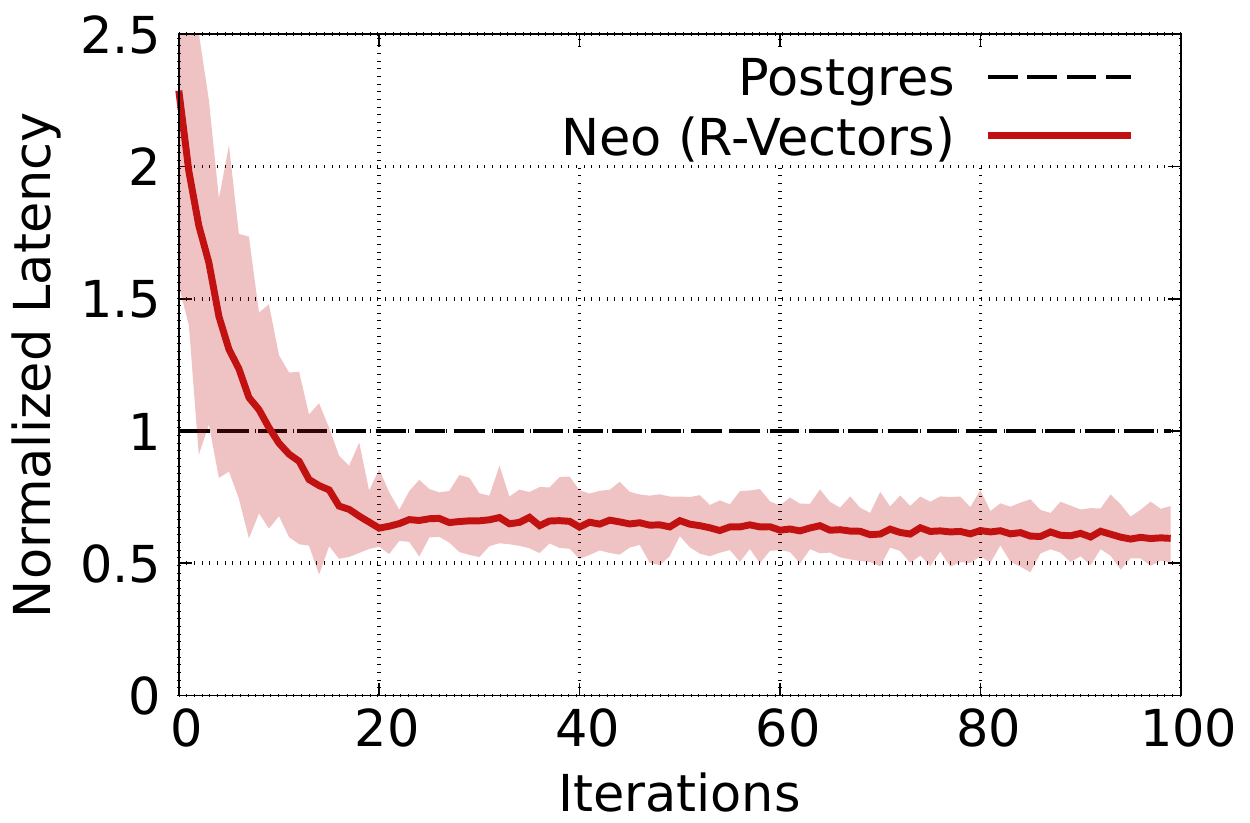}
    & \includegraphics[width=0.22\textwidth]{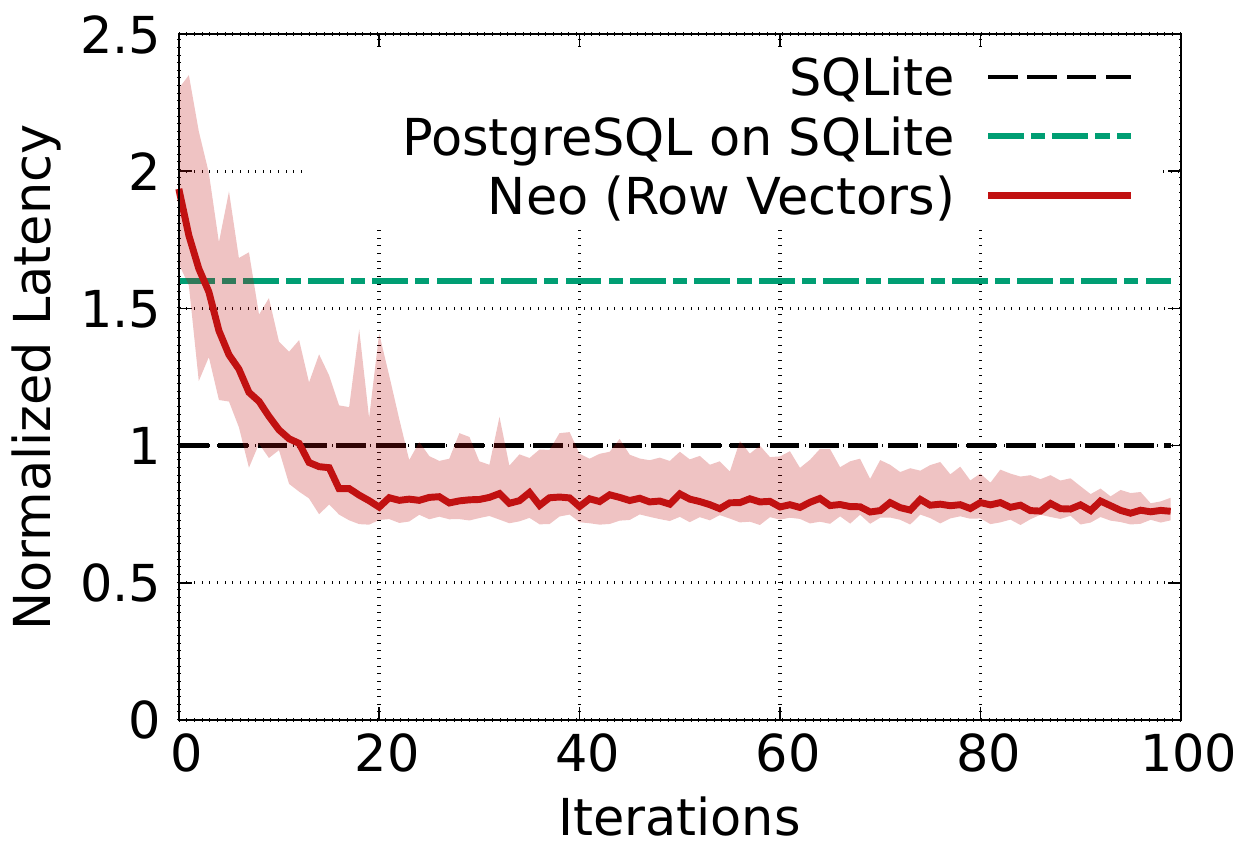}
    & \includegraphics[width=0.22\textwidth]{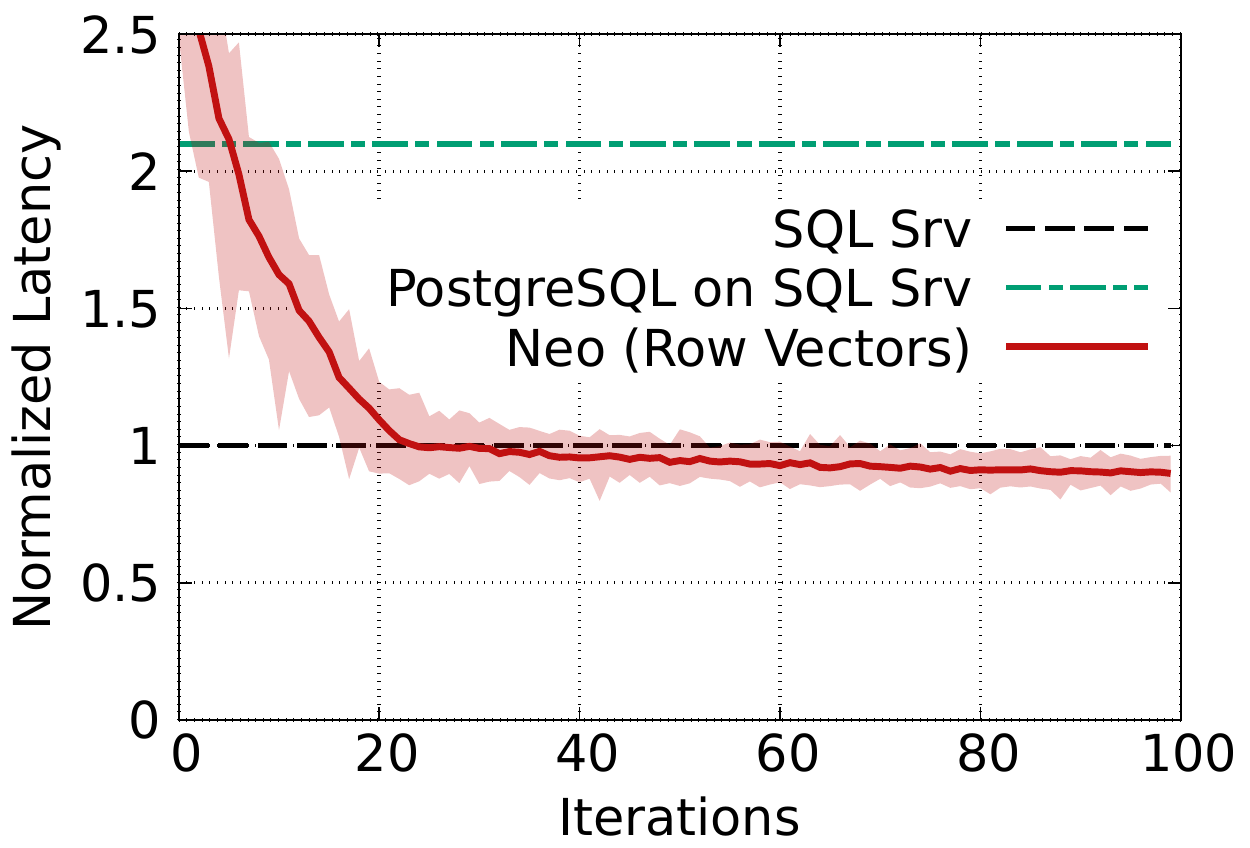}
    & \includegraphics[width=0.22\textwidth]{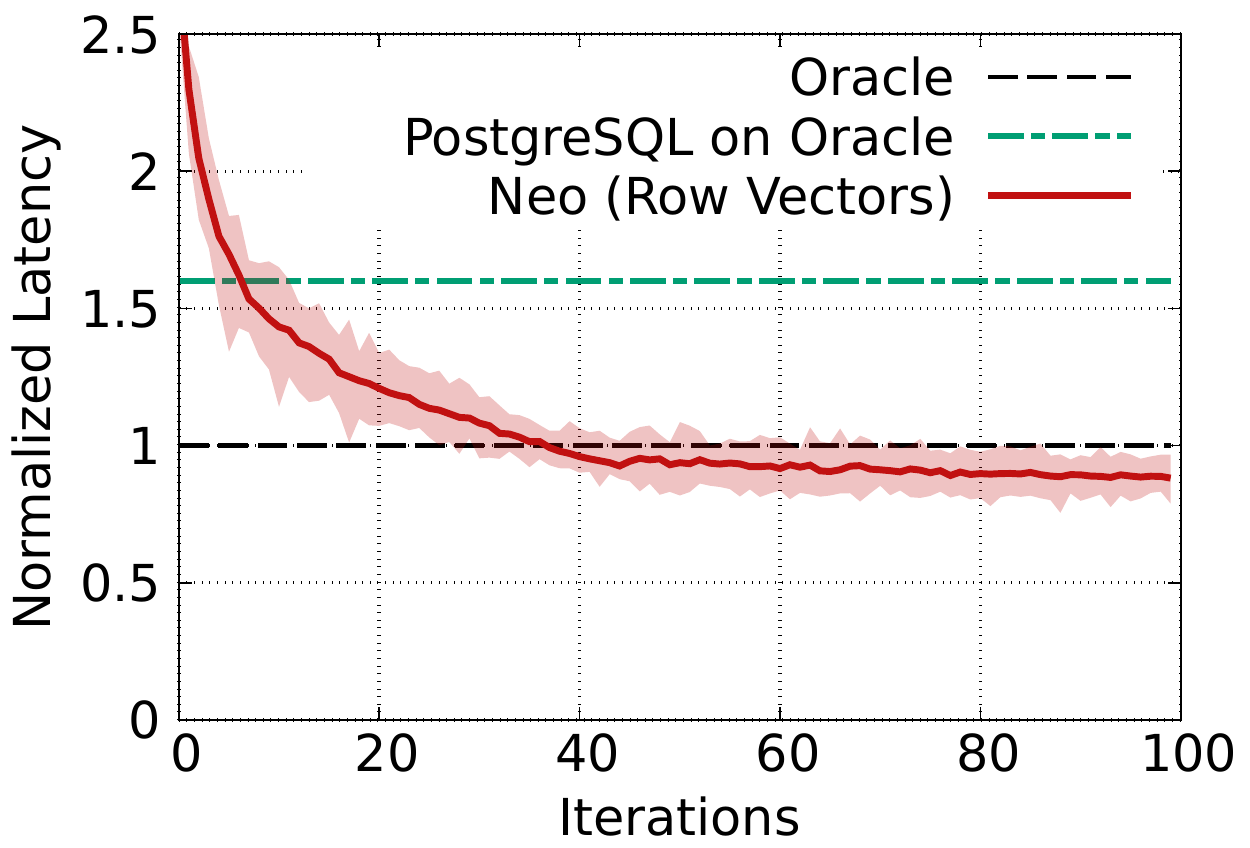} \\
    \rotatebox{90}{\TPC}
    & \includegraphics[width=0.22\textwidth]{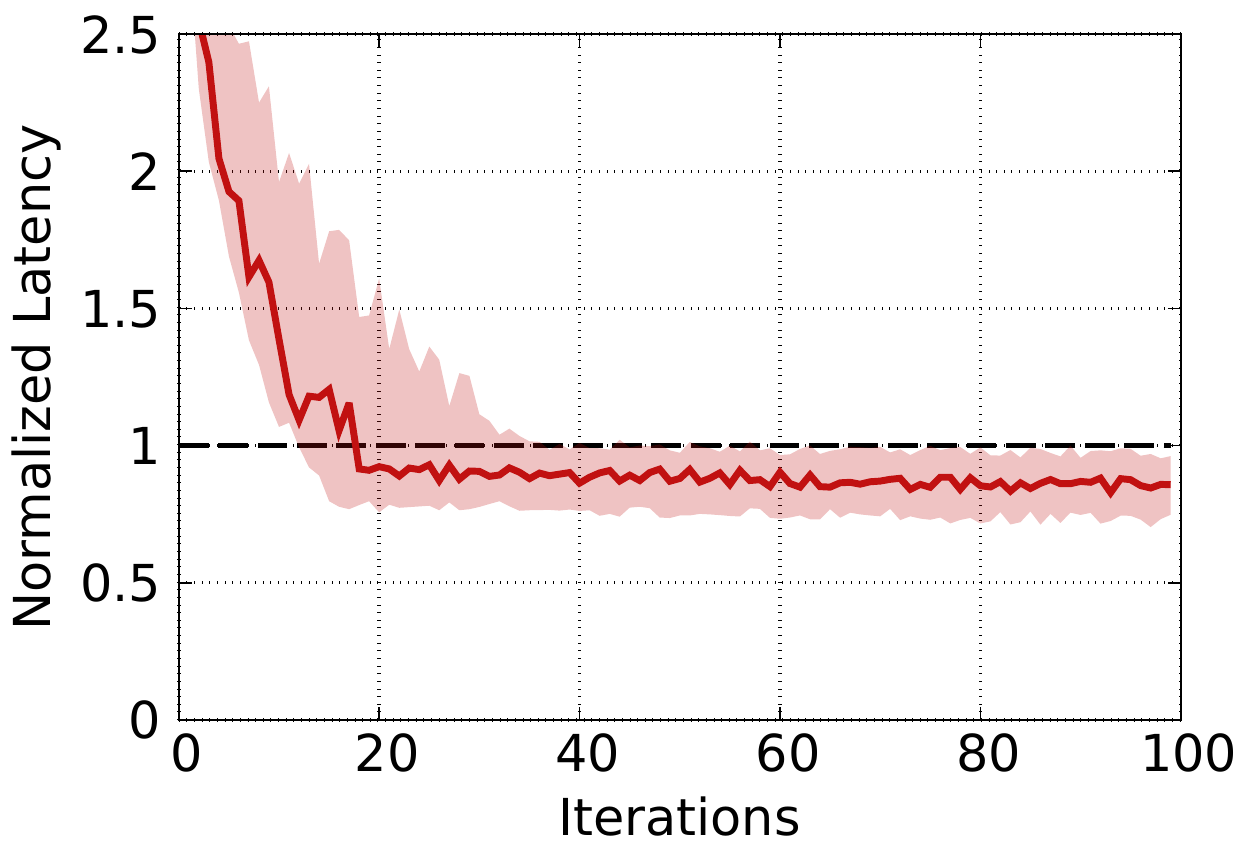}
    & \includegraphics[width=0.22\textwidth]{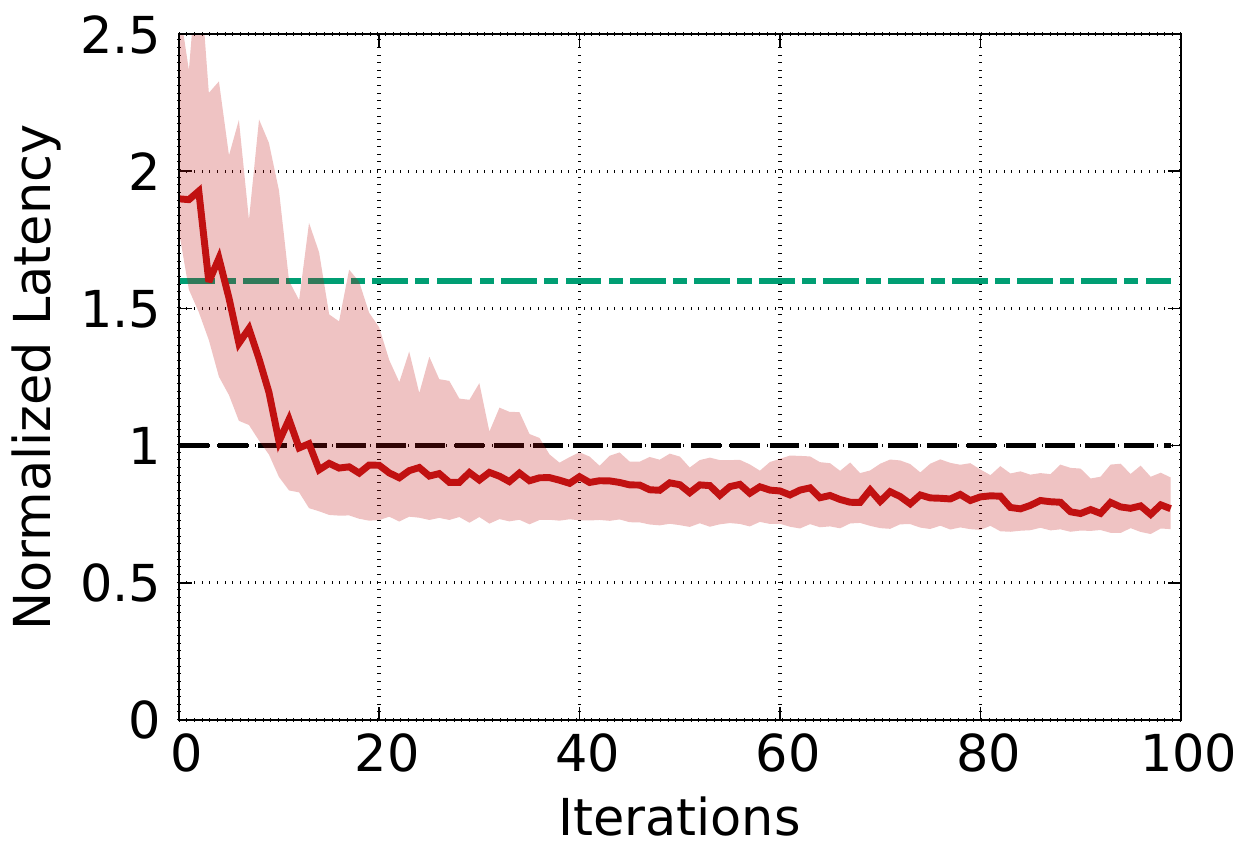}
    & \includegraphics[width=0.22\textwidth]{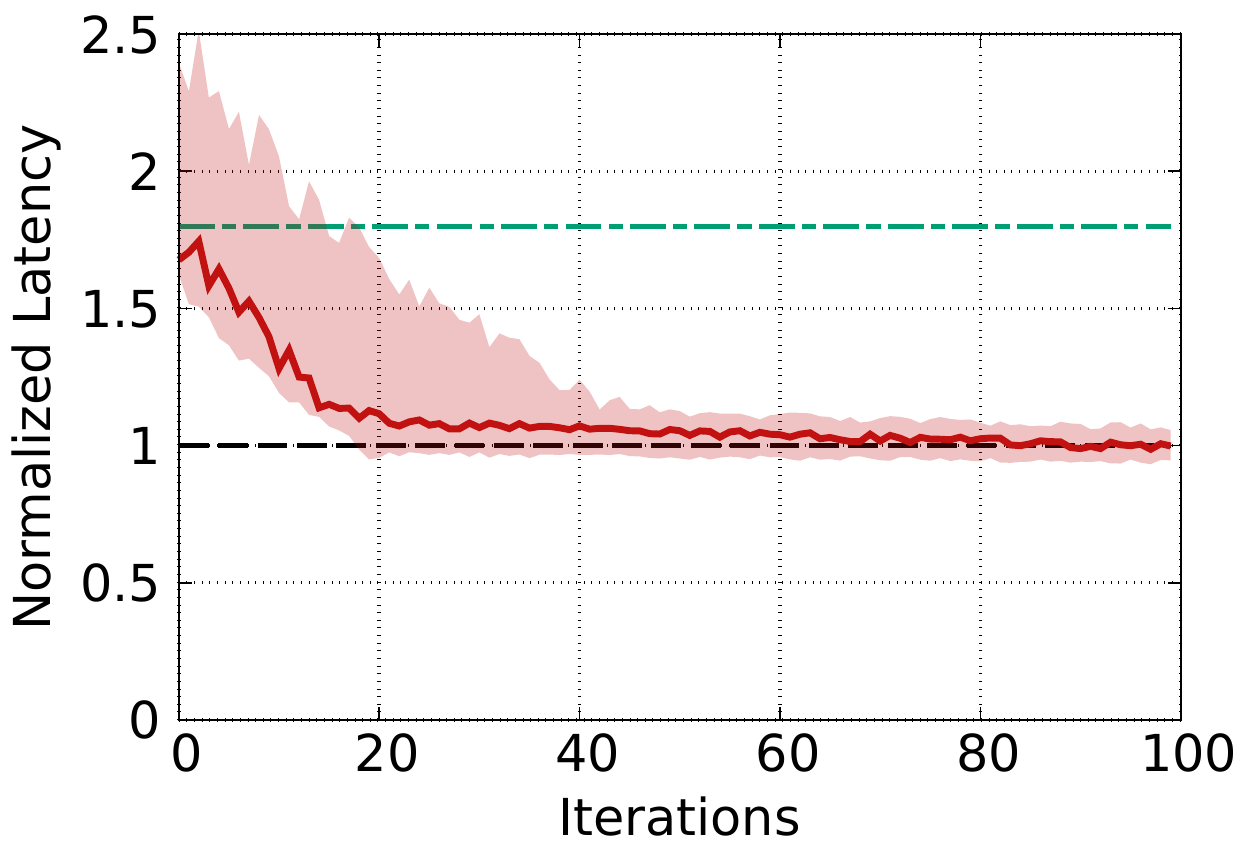}
    & \includegraphics[width=0.22\textwidth]{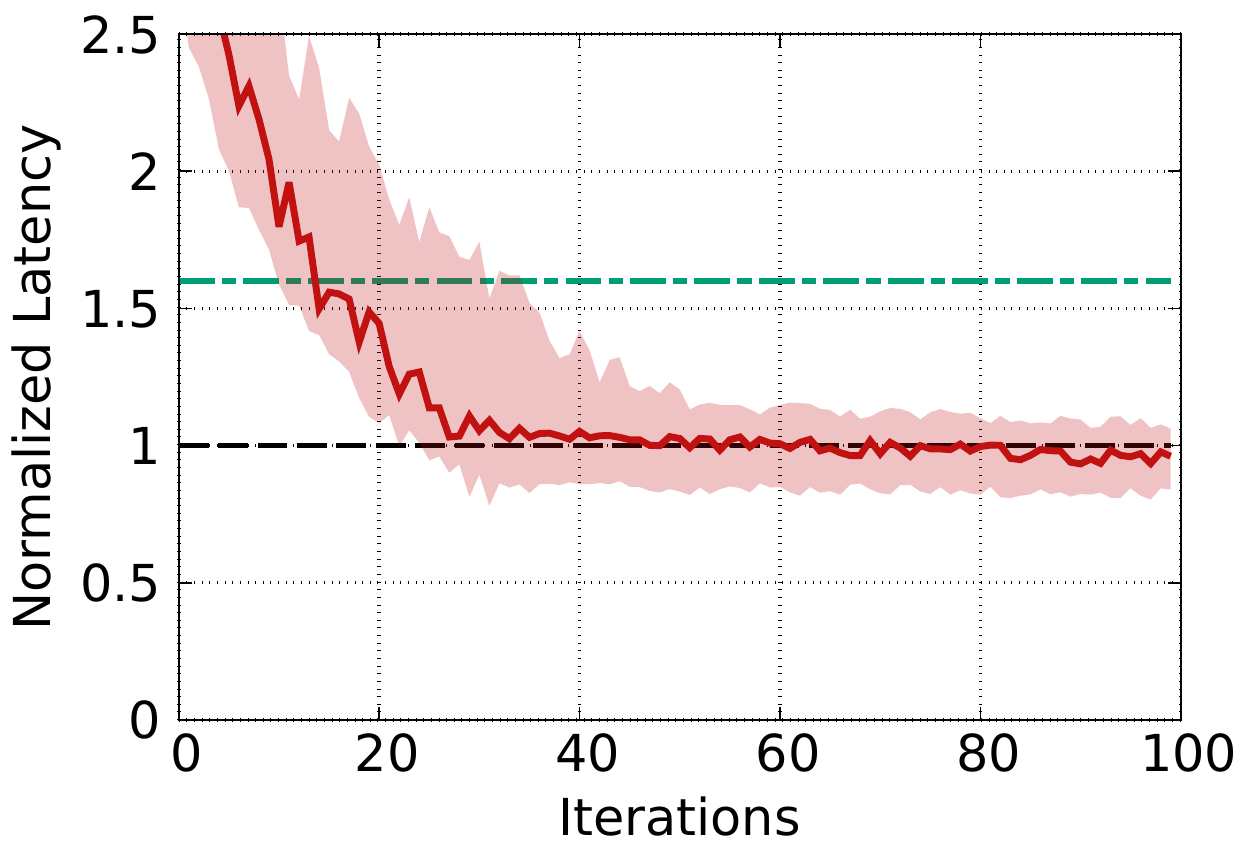} \\
    \rotatebox{90}{\Vertica}
    & \includegraphics[width=0.22\textwidth]{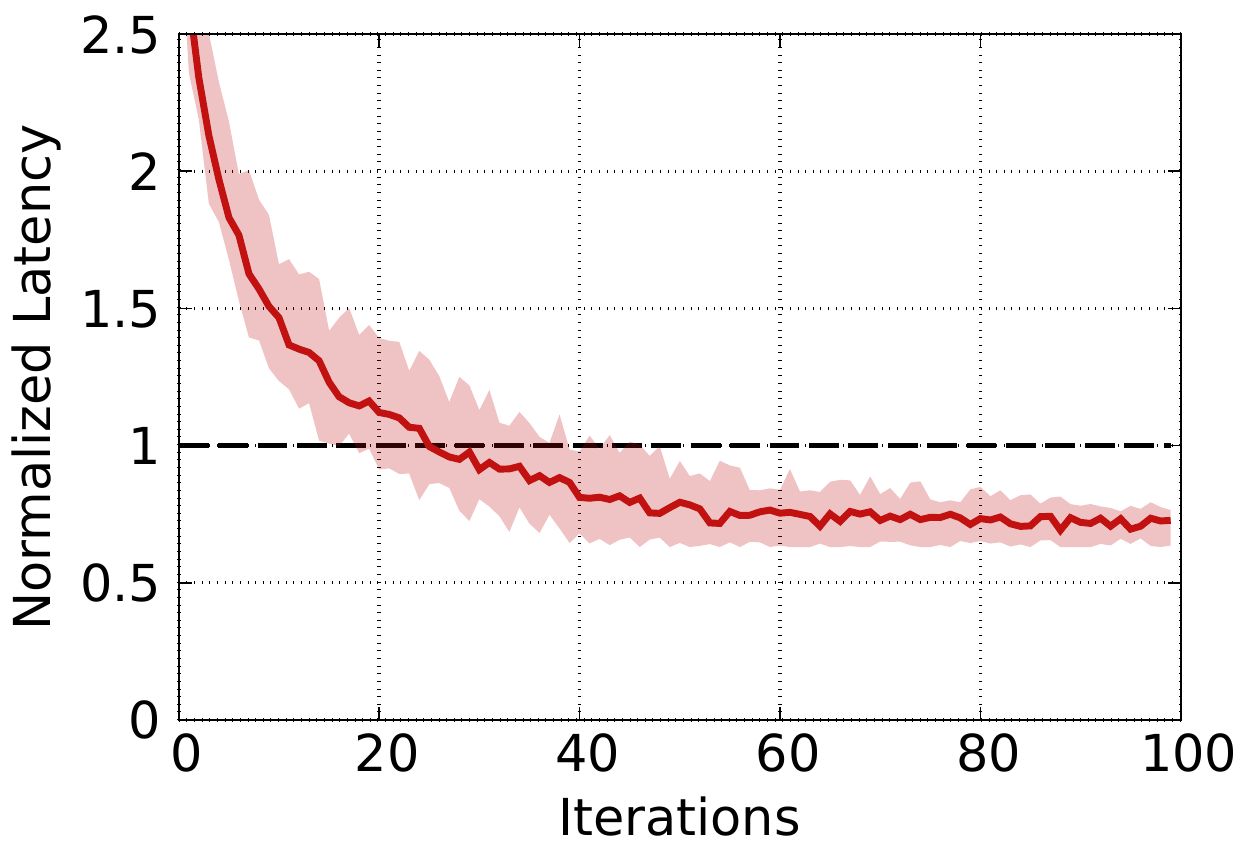}
    & \includegraphics[width=0.22\textwidth]{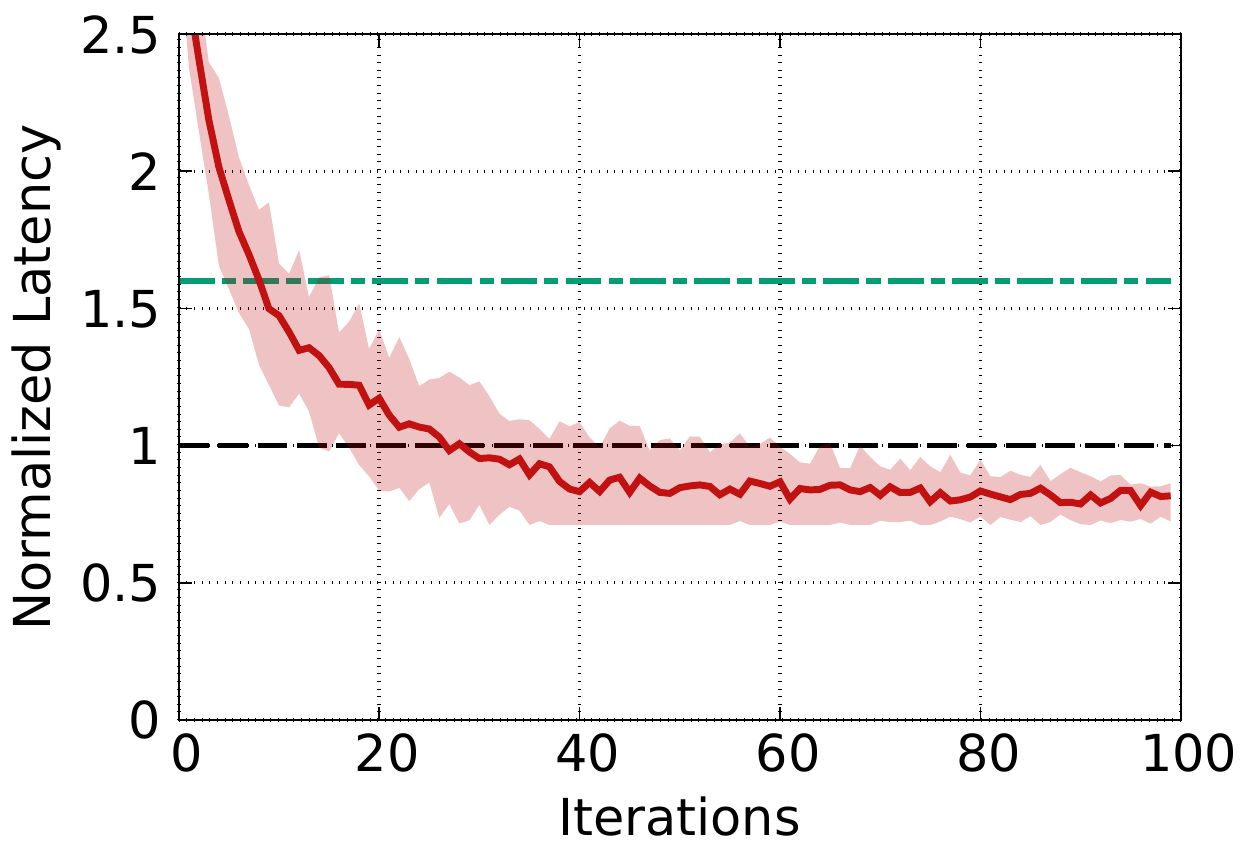}
    & \includegraphics[width=0.22\textwidth]{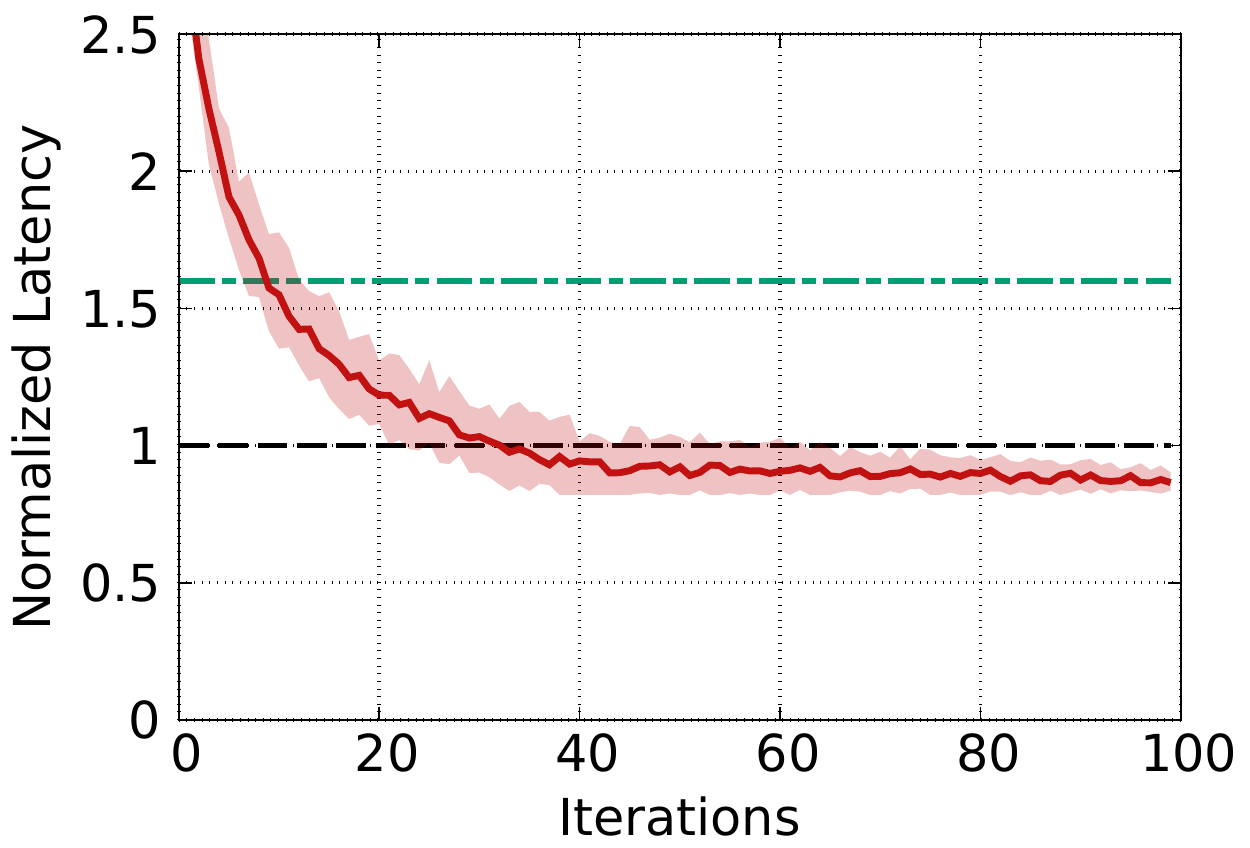}
    & \includegraphics[width=0.22\textwidth]{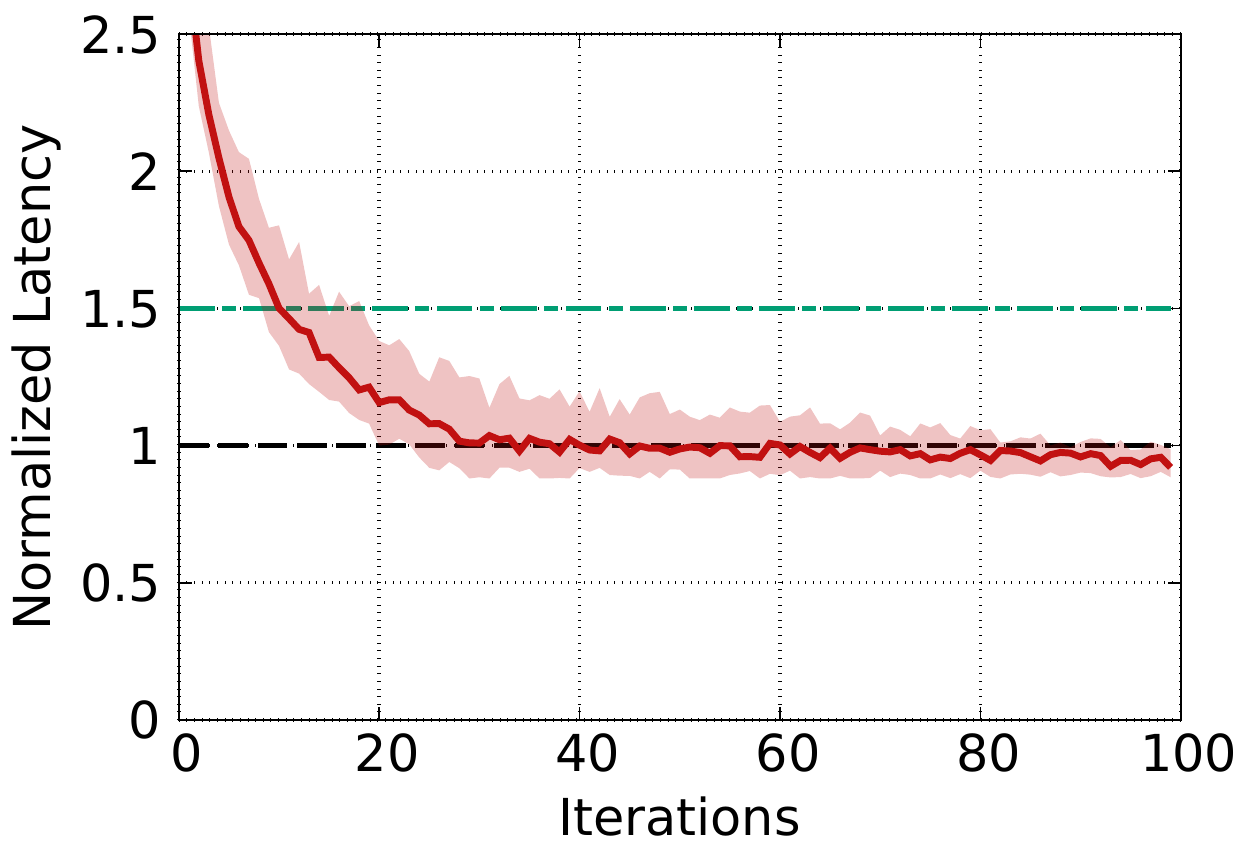}
  \end{tabular}
  \caption{Learning curves with variance. Shaded area spans minimum to maximum across fifty runs with different random seeds. For a plot with all four featurization techniques, please visit: \protect\url{http://rm.cab/l/lc.pdf}}
  \label{fig:curves}
  \vspace{3mm}
\end{figure*}

We measured the relative performance of \system on our entire test suite with respect to the native optimizer (i.e., a performance of $1$ is equivalent to the engine's optimizer), for every episode (a full pass over the set of training queries, i.e., retraining the network from the experience, choosing a plan for each training query, executing that plan, and adding the result to \system's experience) of the 100 complete training episodes of the optimizer. We plot the median value as a solid line, and the minimum and maximum values using the shaded region. For all DBMSes except for \PG, we additionally plot the relative performance of the plans generated by the \PG optimizer when executed on the target engine.

\sparagraph{Convergence} Each figure demonstrates a similar behavior: after the first training iteration, \system's performance is poor (e.g., nearly 2.5 times worse than the native optimizer). Then, for several iterations, the performance of \system sharply improves, until it levels off (converges). We analyze the convergence time specifically in Section~\ref{sec:training_time}. Here, we note that \system is able to improve on the \PG optimizer in as few as 9 training iterations (i.e., the number of training iterations until the median run crosses the line representing \PG). It is not surprising that matching the performance of a commercial optimizer like \MS or \Ora requires significantly more training iterations than for \SQLite, as commercial systems are much more sophisticated.

\sparagraph{Variance} The variance between the different training iterations is small for all workloads, except for the TPC-H dataset. We hypothesize that, with uniform data distributions in TPC-H, the \rv embedding is not as useful, and thus it takes the model longer to adjust accordingly. This behavior is not present in the other two non-synthetic datasets.

%I deleted the SQLite stuff as I do not see a huge difference in the worst case time. 
%Because the \PG optimizer assumes the underlying engine has hash joins, merge joins, etc., the initial workload latency of the \PG execution plans on the SQLite engine is nearly double that of the plans produced by the \SQLite optimizer; \SQLite only supports loop joins, which are especially sensitive to sub-optimal join-orders. 

\subsubsection{Wall-Clock Time}
\label{sec:training_time}

\begin{figure}
    \includegraphics[width=0.46\textwidth]{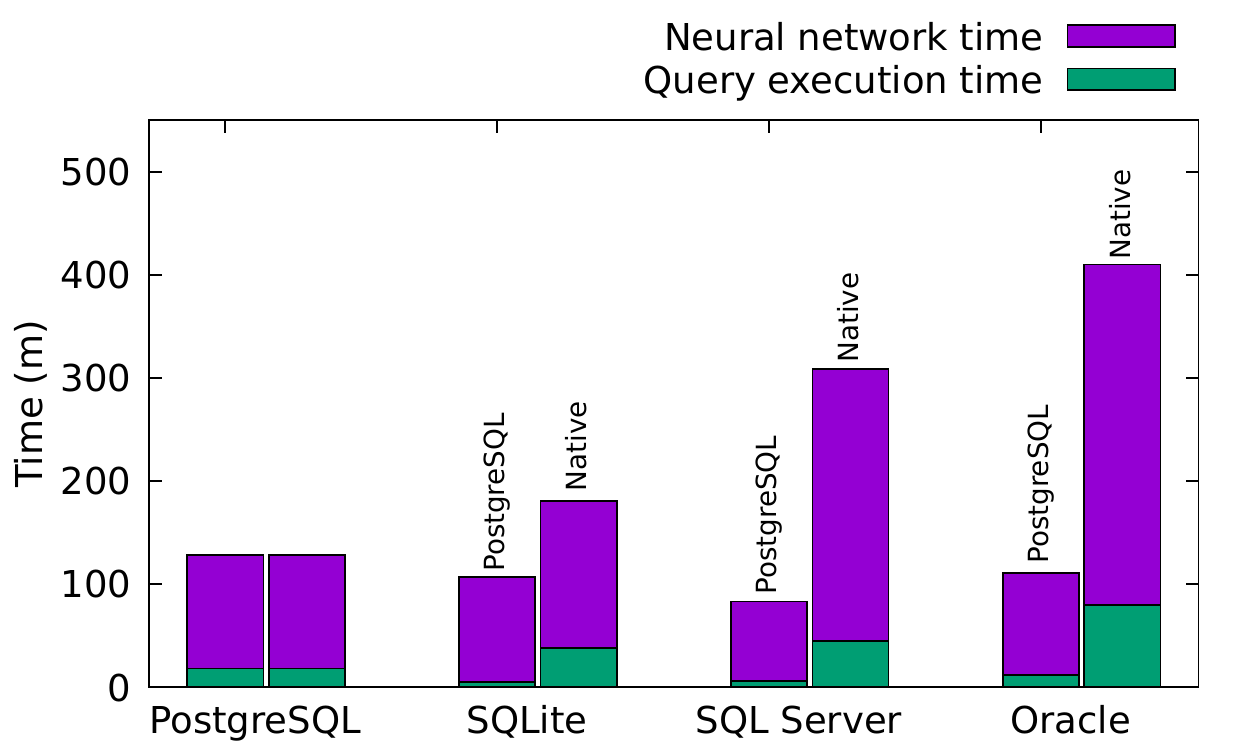}
    \caption{Training time, in minutes, for \system to match the performance of \PG and each native optimizer.}
    \label{fig:walltime}
\end{figure}

%\begin{subfigure}{0.32\textwidth}
%    \centering
%    \includegraphics[width=\textwidth]{experiments/rv_time.pdf}
%    \caption{\rv training time}
%    \label{fig:rv_time}
%  \end{subfigure}

So far, we analyzed how long it took \system to become competitive in terms of \emph{training iterations}; next, we analyze the time it takes for \system to become competitive in terms of \emph{wall-clock time} (real time). We analyzed how long it took for \system to learn a policy that was on-par with (1) the query execution plans produced by \PG, but executed on the target execution engine, and (2) the query plans produced by the native optimizer and executed on the same execution engine. Figure~\ref{fig:walltime} shows the time (in minutes) that it took for \system to reach these two milestones (the left and right bar charts represent milestone (1) and (2), respectively), split into time spent training the neural network and time spent executing queries. Note that the query execution step is parallelized, executing queries on different nodes simultaneously.

Unsurprisingly, it takes longer for \system to become competitive with the more advanced, commercial optimizers. However, for every engine, learning a policy that outperforms the \PG optimizer consistently takes less than two hours. 
Furthermore, \system was able to \emph{match or exceed the performance of every optimizer within half a day}.
Note that this time does not include the time for training the query encoding, which in the case of the {\tt 1-Hot} and {\tt Histogram} are negligible. However, this takes longer for \rv (see Section~\ref{sec:rv_train}).
%For example, for the \JOB dataset (approximately 4GB), the ``no joins'' variant trains in less than 10 minutes, whereas the ``no joins'' variant for the \Vertica dataset (approximately 2TB) requires nearly two hours to train. 
%The ``joins'' (partially denormalized) variant takes significantly longer to train, e.g., from three hours (\JOB) to a full day (27 hours, \Vertica).
%In cases where this cost is prohibitive, the {\tt Histogram} feature provides a very good alternative that still produces a competitive query optimization policy in a reasonable amount of time.

\subsubsection{Is Demonstration Even Necessary?}

Since gathering demonstration data introduces additional complexity to the system, it is natural to ask if demonstration data is necessary at all. Is it possible to learn a good policy starting from zero knowledge? While previous work~\cite{rejoin} showed that an off-the-shelf deep reinforcement learning technique can learn to find query plans \emph{that minimize a cost model} without demonstration data, learning a policy \emph{based on query latency} (i.e., end to end) poses additional difficulties: a bad plan can take hours to execute.
Unfortunately, randomly chosen query plans behave exceptionally poorly. Leis et al. showed that randomly sampled join orderings can result in a 100x to 1000x increase in query execution times for \JOB queries, compared to a reasonable plan~\cite{howgood}, potentially increasing the training time of \system by a similar factor~\cite{cidr_dlqo}. 

We attempted to work around this problem by selecting an ad-hoc query timeout $t$ (e.g., 5 minutes), and terminating query executions when their latencies exceed $t$. However, this technique destroys a good amount of the signal that \system uses to learn: join patterns resulting in a latency of 7 minutes get the same reward as join patterns resulting in a latency of 1 week, and thus \system cannot learn that the join patterns in the 7-minute plan are an improvement over the 1-week plan. 
As a result, even after training for over three weeks, we did not achieve the plan quality that we achieve when bootstrapping the system with the \PG optimizer.

%This makes \system equally likely to further explore the 7 minute plan as the 1 week plan, even though exploring the former is much more likely to be fruitful. As previously noted~\cite{howgood}, the vast majority of query plans are ``generally multiple orders of magnitude more expensive than the cheapest plan'', which means that a \system-style system without demonstration data would be forced to search randomly for a good plan among the multitudes of poor ones. In our experiments, such good plans were never found, even after searching for over three weeks.

\subsection{Robustness}
\label{sec:exp:robustness}

For all experiments thus far, \system was always evaluated over the test dataset, never the training dataset. This clearly demonstrates that \system does generalize to new queries. In this subsection, we study this further by also testing \system's performance for the different featurization techniques, over entirely new queries (i.e., queries invented specifically to exhibit novel behavior), and measuring the sensitivity of chosen query plans to cardinality estimation errors.
%for cases which the cardinality estimates are more or less precise. 

\subsubsection{Featurization}
\label{sec:eval-feature}

  \begin{figure}
    \includegraphics[width=0.46\textwidth]{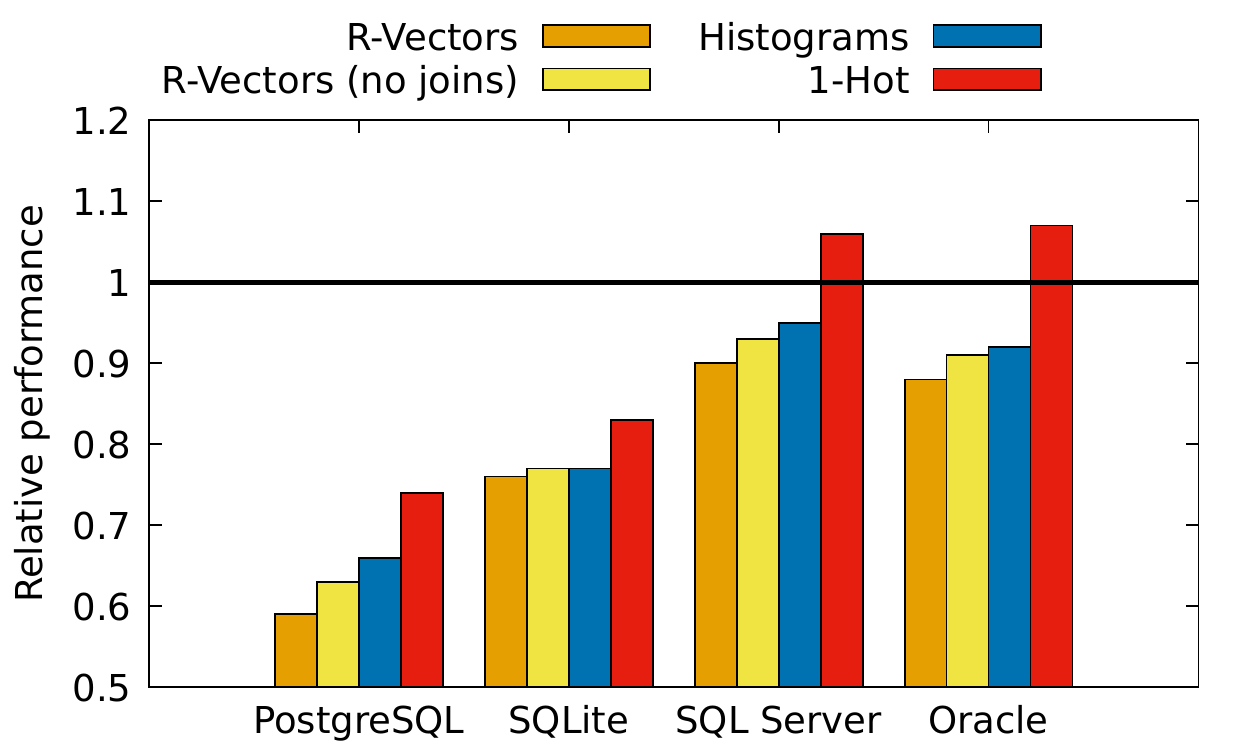}
    \caption{\system's performance using each featurization.}
    \label{fig:trained_by_features}
  \end{figure}
  
%  \end{subfigure}
%    \begin{subfigure}{0.32\textwidth}
%    \centering
%    \includegraphics[width=\textwidth]{experiments/joins_only.pdf}
%    \caption{Join order only, JOB}
%    \label{fig:join_only}
%  \end{subfigure}

Figure~\ref{fig:trained_by_features} shows the performance of \system across all four DBMSes for the \JOB dataset, varying the featurization used. Here, we examine both the regular \rv encoding and a variant of it built without any joins for denormalization (see Section~\ref{sec:pari_features}). As expected, the \oh encoding consistently performs the worst, as the \oh encoding contains no information about predicate cardinality. The \hist encoding, while making naive uniformity assumptions, provides enough information about predicate cardinality to improve \system's performance. In each case, the \rv encoding variants produce the best overall performance, with the ``no joins'' variant lagging slightly behind. This is because the \rv encoding contains significantly more semantic information about the underlying database than the naive histograms (see Section~\ref{sec:pari_features}). The improved performance of \rv compared to the other encoding techniques \emph{demonstrates the benefits of tailoring the feature representation used to the underlying data.}

\subsubsection{On Entirely New Queries}

\begin{figure}
  \centering
  \includegraphics[width=0.46\textwidth]{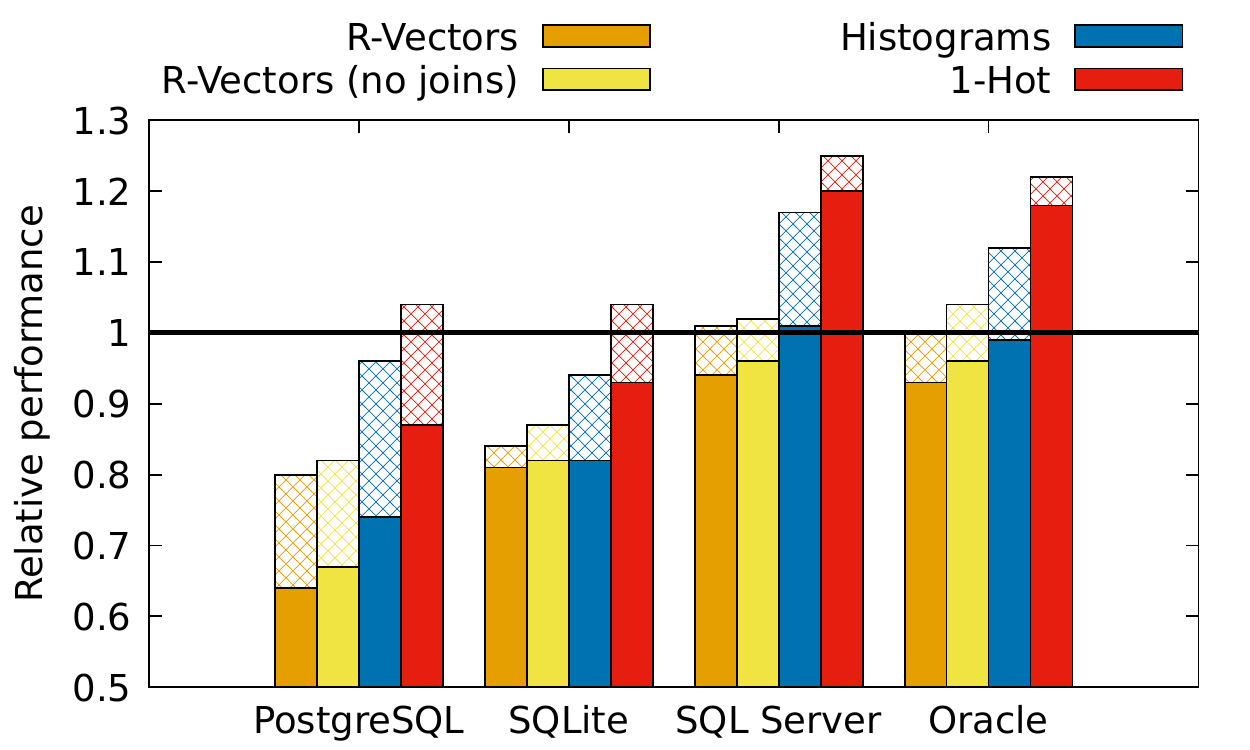}
  \caption{\system's performance on entirely new queries (\EJOB), full bar height. \system's performance after 5 iterations with \EJOB queries, solid bar height.}
  \label{fig:new_queries}
\end{figure}

Previous experiments demonstrated \system's ability to generalize to queries in a randomly-selected, held-out test set drawn from the same workload as the training set. While this shows that \system can handle previously-unseen predicates and modifications to join graphs, it does not necessarily demonstrate that \system will be able to generalize to a completely new query. To test \system's behavior on new queries, we created a set of 24 additional queries\footnote{\url{https://git.io/extended_job}}, which we call \EJOB, that are semantically distinct from the original \JOB workload (no shared predicates or join graphs).

After \system had trained for 100 episodes on the \JOB queries, we evaluated the relative performance of \system on the \EJOB queries. Figure~\ref{fig:new_queries} shows the results: the full height of each bar represents the performance of \system on the unseen queries relative to every other system.  First, we note that with the \rv featurization, \emph{the execution plans chosen for the entirely-unseen queries in the \EJOB dataset still outperformed or matched the native optimizer.} Second, the larger gap between the \rv featurizations and the \hist and \oh featurizations \emph{demonstrates that row vectors are an effective way of capturing information about query predicates that generalizes to entirely new queries.}

\sparagraph{Learning new queries} Since \system is able to progressively learn from each query execution, we also evaluated the performance of \system on the \EJOB queries after just 5 additional training episodes. The solid bars in Figure~\ref{fig:new_queries} show the results. Once \system has seen each new query a handful of times, \system's performance increases quickly, having learned how to handle the new complexities introduced by the previously-unseen queries. Thus, while the performance of \system initially degrades when confronted with new queries, \emph{\system quickly adapts its policy to suit these new queries}. This showcases the potential for a deep-learning powered query optimizer to keep up with changes in real-world query workloads.  %\ma{This is a very nice result. For clarity, it might be better to split Figure 12 into separate figures. Especially 12c is very nice and deserves a proper caption.}
 
%\ma{A question for later: did training on the new queries make Neo forget how to optimize the regular JOB queries? I suspect this ``forgetting'' behavior exists and it might be interesting to think about how to mitigate/avoid it}

\label{sec:expr_robust}
\begin{figure*}
  \centering
  \begin{subfigure}{0.24\textwidth}
    \includegraphics[width=\textwidth]{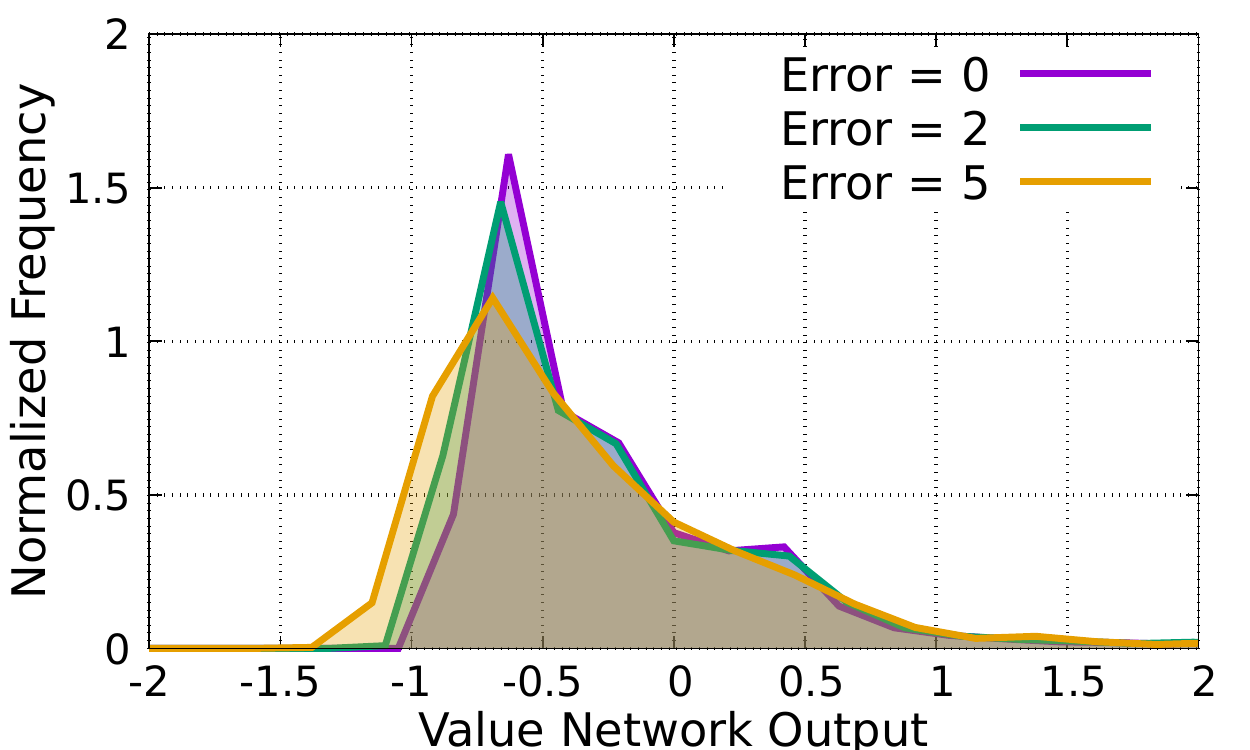}
    \caption{\PG, $\leq$ 3 joins}
    \label{fig:pglte3}
  \end{subfigure}
    \begin{subfigure}{0.24\textwidth}
    \includegraphics[width=\textwidth]{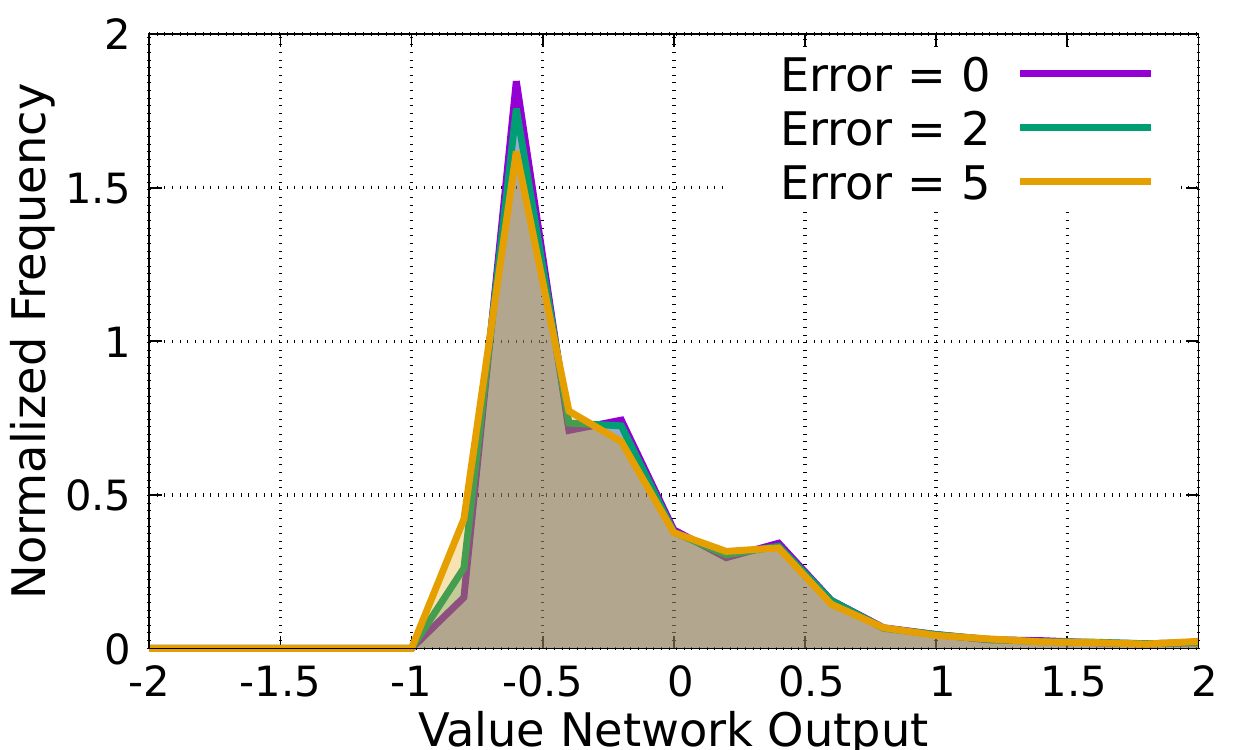}
    \caption{\PG, $>$ 3 joins}
    \label{fig:pggt3}
  \end{subfigure}
  \begin{subfigure}{0.24\textwidth}
    \includegraphics[width=\textwidth]{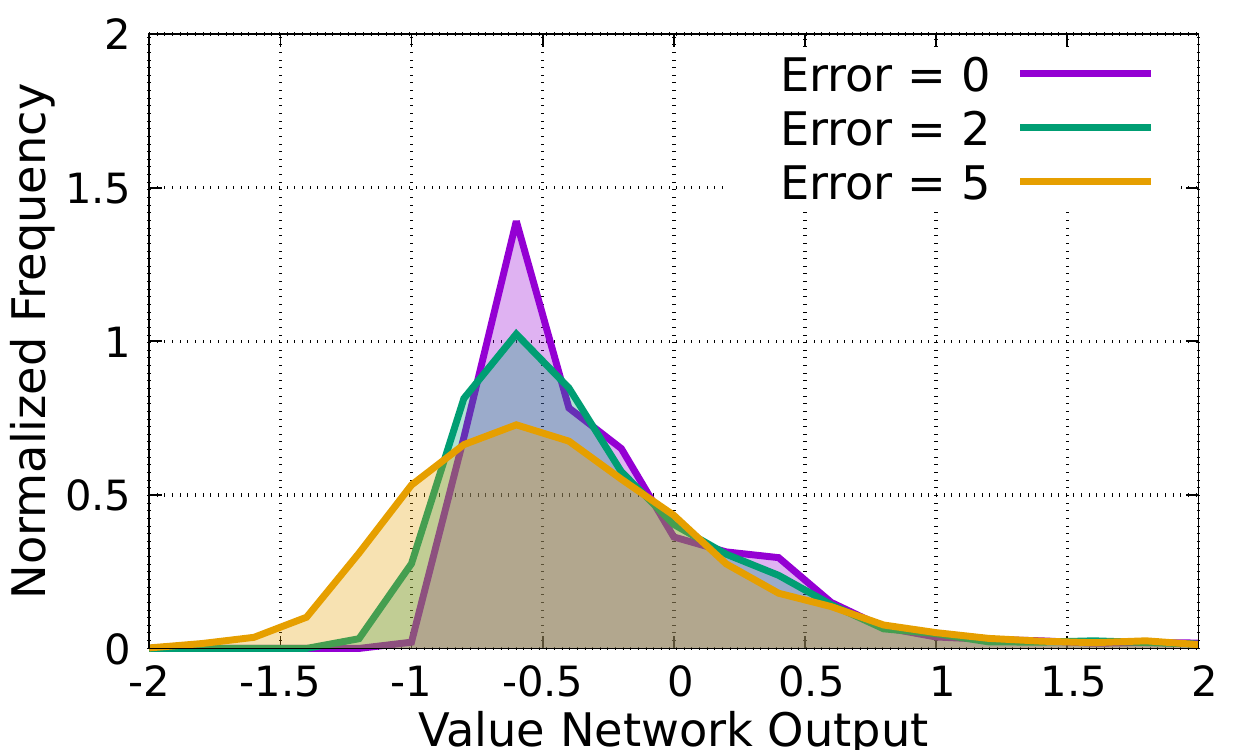}
    \caption{True cardinality, $\leq$ 3 joins}
    \label{fig:truecardlte3}
  \end{subfigure}
  \begin{subfigure}{0.24\textwidth}
    \includegraphics[width=\textwidth]{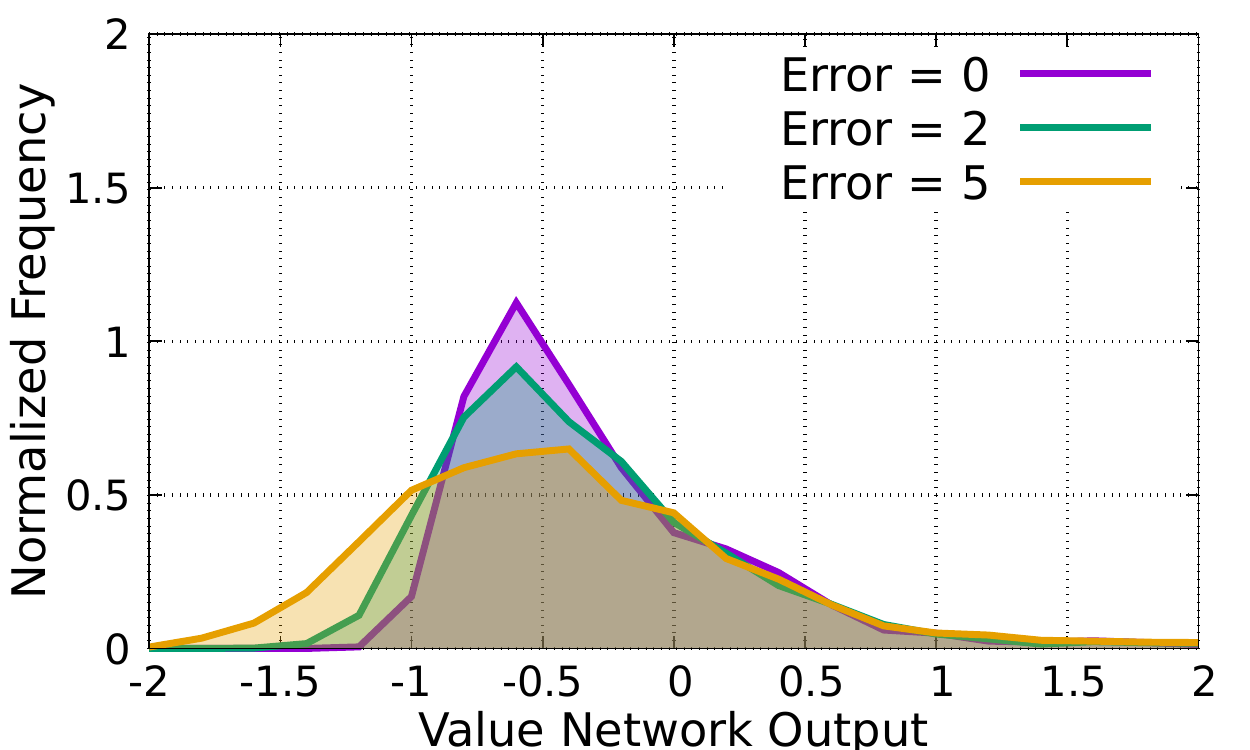}
    \caption{True cardinality, $>$ 3 joins}
    \label{fig:truecardgt3}
  \end{subfigure}
  \caption{Robustness to cardinality estimation errors}
  \label{fig:misc}
\end{figure*}

\subsubsection{Cardinality Estimates}

The strong relationship between cardinality estimation and query optimization is well-studied~\cite{robust_qo,bound_card}. However, query optimizers must take into account that most cardinality estimation methods tend to become significantly less accurate when the number of joins increases~\cite{howgood}. While deep neural networks are generally regraded as black boxes, here we show that \system is capable of learning when to trust cardinality estimates and when to ignore them.

To measure the robustness of \system to cardinality estimation errors, we trained two \system models, with an additional feature at each tree node. The first model received the \PG optimizer's cardinality estimation, and the second model received the true cardinality. We then plotted a histogram of both model's outputs across the \JOB workload when the number of joins was $\leq 3$ and $> 3$, introducing artificial error to the additional features.

For example, Figure~\ref{fig:pglte3} shows the histogram of value network predictions for all states with at most 3 joins. When the error is increased from zero orders of magnitude to two and five orders of magnitude, the variance of the distribution increases: in other words, when the number of joins is at most 3, \system learns a model that varies with the \PG cardinality estimate. However, in Figure~\ref{fig:pggt3}, we see that the distribution of network outputs hardly changes at all when the number of joins is greater than 3: in other words, when the number of joins is greater than 3, \system learns to ignore the \PG cardinality estimates all together.

On the other hand, Figure~\ref{fig:truecardlte3} and Figure~\ref{fig:truecardgt3} show that when \system's value model is trained with true cardinalities as inputs, \system learns a model that varies its prediction with the cardinality regardless of the number of joins. In other words, when provided with true cardinalities, \system learns to rely on the cardinality information irrespective of the number of joins. Thus, we conclude that \system is able to learn which input features are reliable, even when the reliability of those features varies with the number of joins.

\subsubsection{Per Query Performance}
\label{sec:expr_opt}
\begin{figure*}
  \includegraphics[width=\textwidth]{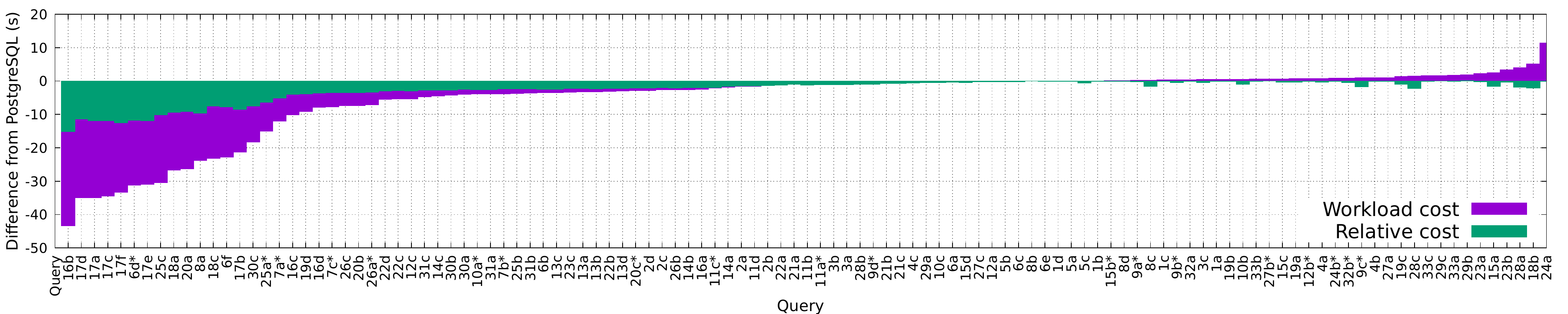}
  \caption{Workload cost vs. relative cost for \JOB queries between \system and \PG (lower is better)}
  \label{fig:loss}
\end{figure*}

Finally, we analyzed the per-query performance of \system  (as opposed to the workload performance).
The absolute performance improvement (or regression) in seconds for each query of the \JOB workload between the \system and \PG plans are shown in Figure~\ref{fig:loss}, in purple. 
As it can be seen, \system is able to significantly improve the execution time of many queries up to 40 seconds, but also worsens the execution time of a few of queries e.g., query 24a becomes 8.5 seconds slower.

However, in contrast to a traditional optimizer, in \system we can easily change the optimization goal. 
So far, we always aimed to optimize the total workload cost, i.e., the total latency across all queries. 
However, we can also change the optimization goal to optimize for the relative improvement per query (green bars in Figure~\ref{fig:loss}). 
This implicitly penalizes changes in the query performance from the baseline (e.g., \PG). 
 When trained with this cost function, the total workload time is still accelerated (by 289 seconds, as opposed to nearly 500 seconds), \emph{and} all but one query\footnote{Query 29b regresses by 43 milliseconds.} sees improved performance from the \PG baseline. Thus, we conclude that \emph{\system responds to different optimization goals, allowing it to be customized for different scenarios and for the user's needs.}

It is possible that \system's loss function could be further customized to weigh queries differently depending on their importance to the user, i.e. query priority. It may also be possible to build an optimizer that is directly aware of service-level agreements (SLAs). We leave such investigations to future work.

%\subsection{Only join order enumeration}
%For some applications, the variance in workload latency provided by the end-to-end learned optimizer may be prohibitive (for example, for real-time analytics). In these cases, \system can be used only to choose a join ordering, leaving operator and index selection to the underlying DBMS. Note that unlike previous join-order-only systems~\cite{rejoin, sanjay_wat}, \system is still trained on the actual query latency, and not on a cost model. Figure~\ref{fig:join_only} shows the learning curve on \PG for the \JOB benchmark when only join ordering is performed for each featurization. While the final performance is somewhat degraded from the end-to-end case (73\% vs. 60\%) (which is unsurprising, given the high importance of join order selection~\cite{howgood}), the variance is significantly decreased (approximately half).
%\tim{if you need space, we can delete this experiment. Actually, I would delete it for sure. It is not that exciting and does}

\subsection{Search}
\label{sec:expr_search}
 \begin{figure}
    \includegraphics[width=0.47\textwidth]{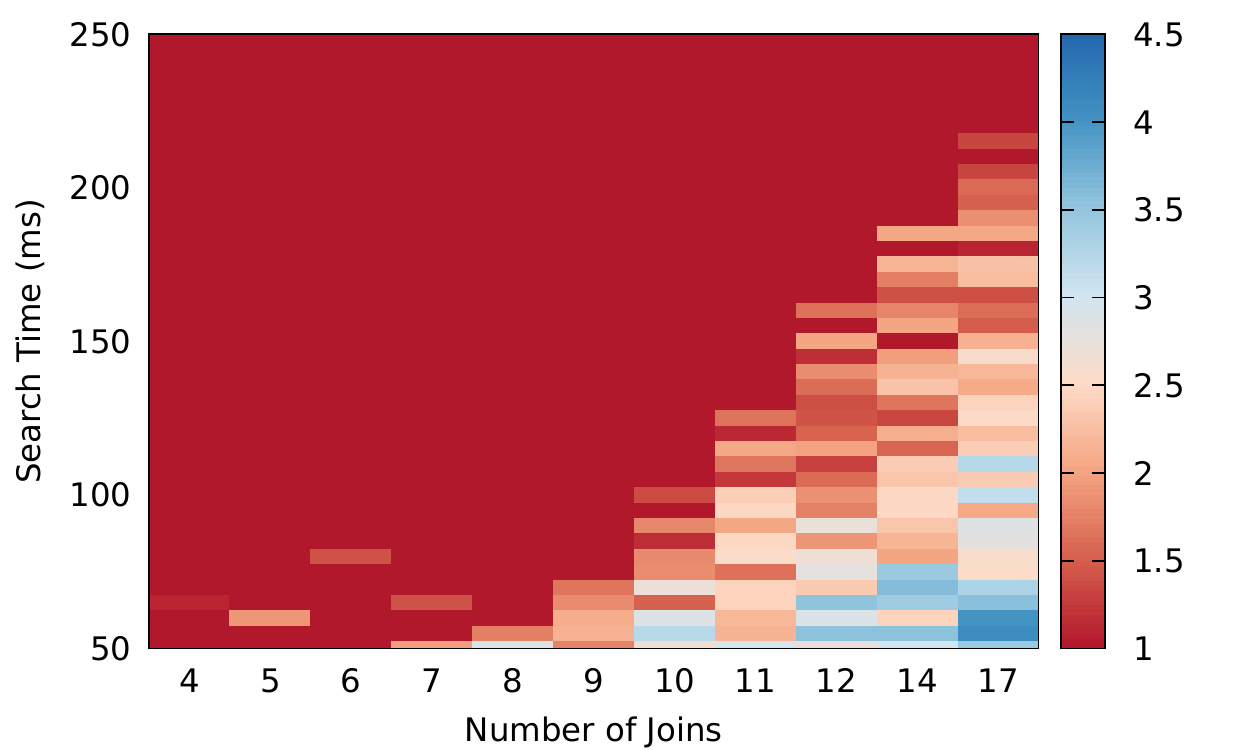}
    \caption{Search time vs. performance, grouped by number of joins}
    \label{fig:search_by_joins}
\end{figure}

\system uses the trained value network to search for query plans until a fixed-time cutoff (see Section~\ref{sec:search}). Figure~\ref{fig:search_by_joins} shows how the performance of a query with a particular number of joins (selected randomly from the \JOB dataset, executed on \PG) varies as the search time is changed (previous experiments used a fixed cutoff of 250ms). Note that the x-axis skips some values, e.g. the \JOB dataset has no queries with 13 joins. Here, query performance is given relative to the best observed performance. For example, when the number of joins is 10, \system found the best-observed plan whenever the cutoff time was greater than 120ms. We also tested significantly extending the search time (to 5 minutes), and found that such an extension did not change query performance regardless of the number of joins in the query (up to 17 in the \JOB dataset).

The relationship between the number of joins and sensitivity to search time is unsurprising: queries with more joins have a larger search space, and thus require more time to optimize. While 250ms to optimize a query with 17 joins is acceptable in many scenarios, other options~\cite{quickpick} may be more desirable when this is not the case.

\subsection{Row vector training time}
\label{sec:rv_train}

\begin{figure}
  \centering
    \includegraphics[width=0.46\textwidth]{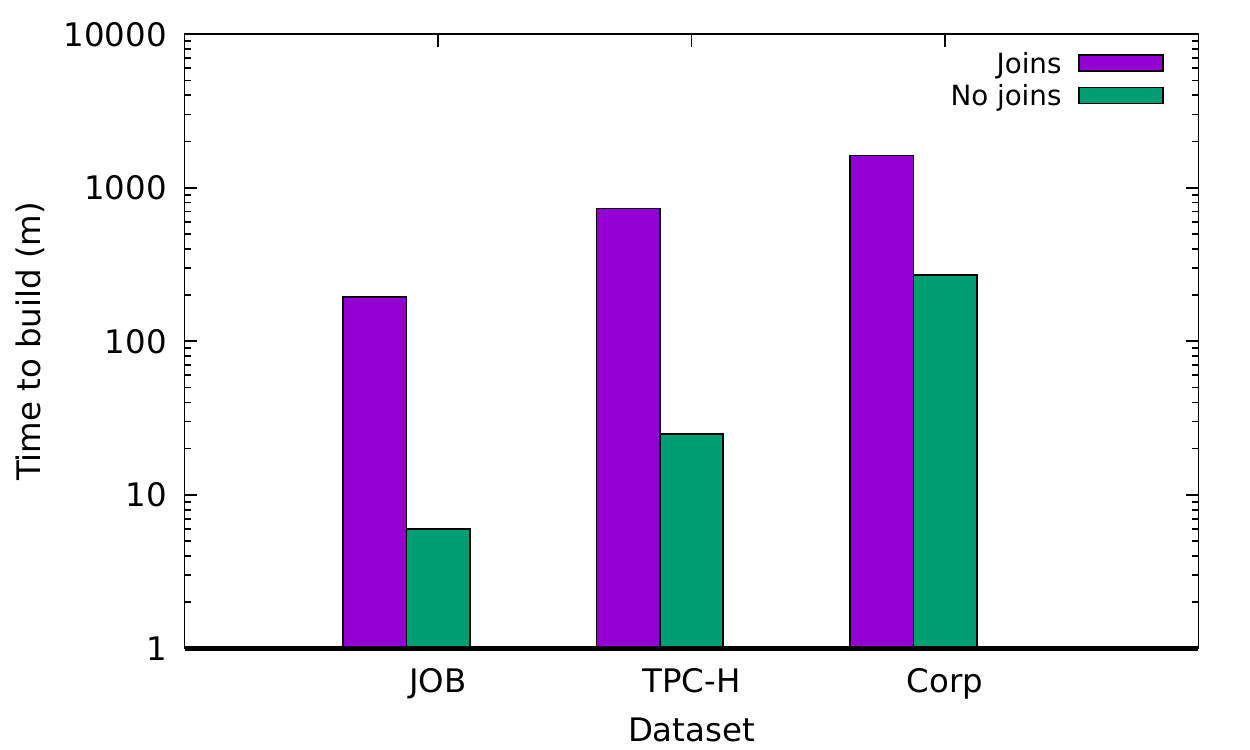}
    \caption{Row vector training time}
    \label{fig:rv_time}
\end{figure}

Here, we analyze the time it takes to build the \rv representation. Our implementation uses the open source \texttt{gensim} package~\cite{gensim}, with no additional optimizations. Figure~\ref{fig:rv_time} shows the time it takes to train row vectors on each dataset, for both the ``joins'' (partially denormalized) and ``no joins'' (normalized) variants, as described in Section~\ref{sec:pari_features}. The time to train a row embedding model is proportional to the size of the dataset. For the \JOB dataset (approximately 4GB), the ``no joins'' variant trains in less than 10 minutes, whereas the ``no joins'' variant for the \Vertica dataset (approximately 2TB) requires nearly two hours to train. The ``joins'' (partially denormalized) variant takes significantly longer to train, e.g. three hours (\JOB) to a full day (27 hours, \Vertica).

Building either variant of row vectors may be prohibitive in some cases. However, experimentally we found that, compared to \hist, the ``joins'' variant on average resulted in 5\% faster query processing times and that the ``no joins'' variant on average resulted in 3\% faster query processing times (e.g., Figure~\ref{fig:trained}). Depending on the multiprocessing level, query arrival rate, etc., row vectors may ``pay for themselves'' very quickly: for example, the training time for building the ``joins'' variant on the \Vertica dataset is ``paid for'' after 540 hours of query processing, since the row vectors speed up query processing by 5\% and require 27 hours to train. As the corporation constantly executes 8 queries simultaneously, this amounts to just three days. The ``no joins'' variant (improves performance by 3\%, takes 217 minutes to train) is ``paid for'' after just 15 hours.

We do not analyze the behavior of row vectors on a changing database. It is possible that, depending on the database, row vectors quickly become ``stale'' (the data distribution shifts quickly), or remain relevant for long periods of time (the data distribution shifts slowly). New techniques~\cite{retrofit_wv, refine_wv} suggest that retraining word vector models when the underlying data has changed can be done quickly, but we leave investigating these methods to future work.

%\subsection{Other experts}
%A natural next step for \system is to test using other optimizers as an expert, instead of the open-source one provided with \PG. For example, using a commercial database system as an initial expert might provide substantially faster convergence, or even a better final result. Alternatively, using a simple, Selinger-style~\cite{systemr} optimizer may prove effective, alleviating the need for even the complexities of the \PG optimizer.

%This work does not evaluate using commercial systems as an initial expert for \system because (1) the closed-source nature of these commercial systems makes extracting information about execution plans are built more difficult, and (2) depending on the existence of complex commercial systems \emph{a priori} defeats a major purpose of \system, i.e. to avoid the need to continually develop and maintain such complex systems in the first place. Regardless, assuming gathering the necessary information about execution plans is possible, we plan to test \system's performance when bootstrapped on commerical systems in future work.

%\begin{figure*}
%  \centering
%  \begin{subfigure}{0.32\textwidth}
%    \includegraphics[width=\textwidth]{experiments/search.pdf}
%    \caption{Search time vs. workload performance}
%    \label{fig:search}
%  \end{subfigure}
%  \caption{}
%  \label{fig:misc}
%\end{figure*}

%%% Local Variables:
%%% mode: latex
%%% TeX-master: "main"
%%% End:

%% file: related.tex
\section{Related work}
\label{sec:related}

% \sparagraph{Query Optimization}
The relational query optimization problem has been around for more than forty years and is one of the most studied problems in database management systems~\cite{systemr, overview98}.
% Variants of cost- and rule-based approaches from the early IBM System R era are still being widely used by today's mature industrial systems \cite{overview98}.
Yet, query optimization is still an unsolved problem~\cite{qo_unsolved}, especially due to the difficulty of accurately estimating cardinalities~\cite{howgood, job2}. 
% While there has been much research on statistical techniques (e.g., histograms \cite{histograms}, selectivity estimation \cite{selectivity}), typical simplifying assumptions such as uniformity, independence, and principle of inclusion do not hold in real-world workloads, leading to unexpected query latencies \cite{job2}.
% IBM DB2's LEO optimizer was the first to introduce the idea of a query optimizer that learns from its mistakes \cite{leo}. LEO continuously monitors query execution and compares actual run-time cardinalities to optimizer's estimates in order to identify column correlation errors and correct its cost model accordingly. In follow-up work, CORDS proactively discovers such correlations between any two columns using data samples in advance of query execution \cite{cords}.
IBM DB2's LEO optimizer was the first to introduce the idea of a query optimizer that learns from its mistakes~\cite{leo}. In follow-up work, CORDS proactively discovered correlations between any two columns using data samples in advance of query execution~\cite{cords}.
%Tlo

Recent progress in machine learning has led to new ideas for learning-based approaches, especially deep learning~\cite{dbml}, to optimizing query run time. % query optimization~\cite{dbml}.
%
% Previous work also used reinforcement learning techniques in query processing.
For example, recent work ~\cite{adaptive_qp_rl, cuttlefish} showed how to exploit reinforcement learning for Eddies-style, fine-grained adaptive query processing. More recently, Trummer et al. have proposed the SkinnerDB system, based on the idea of using regret bound as a quality measure while using reinforcement learning for dynamically improving the execution of an individual query in an adaptive query processing system~\cite{skinnerdb}.
Ortiz et al. analyzed how state representations affect query optimization when using reinforcement learning~\cite{qo_state_rep}.
QuickSel offered using query-driven mixture models as an alternative to using histograms and samples for adaptive selectivity learning \cite{quicksel}. 
Kipf et al. and Liu et al. proposed a deep learning approach to cardinality estimation, specifically designed to capture join-crossing correlations and 0-tuple situations (i.e., empty base table samples)~\cite{deep_card_est, deep_card_est2}.
%
% Similar in spirit to our row embeddings, learned vector embeddings have also found use in representing other relational database features (e.g., query operators \cite{op_embed}, SQL queries \cite{sql_embed}, column relationships in tables \cite{column_embed}, and relational entities~\cite{deep_entity}). 
%
The closest work to ours is DQ~\cite{sanjay_wat},
%and ReJOIN~\cite{rejoin},
which proposed a learning based approach exclusively for join ordering, and only for a given cost model. 
The key contribution of our paper over all of these previous approaches is that it provides an end-to-end, continuously learning solution to the database query optimization problem. Our solution does not rely on any hand-crafted cost model or data distribution assumptions.
%, or cost models, or cardinality estimation.

% Previous work also used reinforcement learning (RL) techniques in query processing. For example, Tzoumas et al. showed how to exploit RL for Eddies-style, fine-grained adaptive query processing \cite{adaptive_qp_rl}. More recently, Trummer et al. have proposed the SkinnerDB system based on the idea of using regret bound as a quality measure in using RL for dynamically improving the execution of an individual query \cite{skinnerdb}.

This paper builds on recent progress from our own team. ReJOIN~\cite{rejoin} proposed a deep reinforcement learning approach for join order enumeration \cite{rejoin}, which was generalized into a broader vision for designing an end-to-end learning-based query optimizer in~\cite{cidr_dlqo}. Decima~\cite{decima} proposed a reinforcement learning-based scheduler, which processes query plans via a graph neural network to learn workload-specific scheduling policies that minimize query latency.  SageDB~\cite{sagedb} laid out a vision towards building a new type of data processing system which will  replace every component of a database system, including the query optimizer, with learned components, thereby gaining the capability to best specialize itself for every use case. This paper is one of the first steps to realizing this overall vision.

% \nt{Remove the RL paragraph below?}

% \sparagraph{Reinforcement learning}
% RL has a long history of development~\cite{rl_book}. Recently, DQN~\cite{dqn} first leverages modern neural networks to achieve super-human performance in Atari video games. The heart of RL is a systematic trial-and-error procedure guided by direct interactions with the environment. Such approach is general and model-free, which does not rely on hard assumptions on the application details. As a result, a surge of recent work has applied RL to a wide range of applications—e.g., robotic control~\cite{rl-robotics-survey}, industrial manufacturing~\cite{rl-manufacturing}, strategic games like Dota~\cite{openai-five} and Starcraft~\cite{alphastar} and data center job scheduling~\cite{decima}. Among the most frontier work, AlphaZero~\cite{alphazero} successfully explores exponentially large policy space in complex board games like Go, by integrating RL with general purpose tree search, similar to the core technique in our design (Section \ref{sec:lo}). \hongzi{citation in hongzi.bib}

%%% Local Variables:
%%% mode: latex
%%% TeX-master: "main"
%%% End:

%% file: conclusions.tex
\section{Conclusions}
\label{sec:conclusions}

This paper presents \system, the first end-to-end learning optimizer that generates highly efficient query execution plans using deep neural networks. \system iteratively improves its performance through a combination of reinforcement learning and a search strategy. On four database systems and three query datasets, \system consistently outperforms or matches existing commercial query optimizers (e.g., Oracle's and Microsoft's) which have been tuned over decades. 

In the future, we plan to investigate various methods for generalizing a learned model to unseen schemas (using e.g. transfer learning~\cite{transfer}). We also intend to further optimize our row vector encoding technique. Finally, we are interested in measuring the performance of \system when bootstrapping from both more primitive and advanced commercial optimizers. Using a commercial database system as an initial expert might provide substantially faster convergence, or even a better final result (although this would introduce a dependency on a complex, hand-engineered optimizer, defeating a major benefit of \system). Alternatively, using a simple, Selinger-style~\cite{systemr} optimizer may prove effective, alleviating the need for even the complexities of the \PG optimizer.

%% file: apx_model.tex
\section{neural network model}
\label{apx:model}
In this appendix, we present a formal specification of \system's neural network model (the value network). An intuitive description is provided in Section~\ref{sec:value_network}.

Let the query-level information vector for an execution plan $P$ for a query $Q(P)$ be $V(Q(P))$. Let $F(P)$ be the set of root nodes in the (forest) $P$. We define $L(x)$ and $R(x)$ as the left and right children of a node, respectively. Let $V(x)$ be the vectorized representation of the tree node $x$. We denote all $r \in F(P)$ as the tuple $(r, L(r), R(r))$.

The query-level information $V(Q(P))$ is initially passed through a set of fully connected layers (see Figure~\ref{fig:network}) of monotonically decreasing size. After the final fully connected layer, the resulting vector $\vec{g}$ is combined with each tree node to form an \emph{augmented forest} $F^\prime(P)$. Intuitively, this augmented forest is created by appending $\vec{g}$ to each tree node. Formally, we define $A(r)$ as the augmenting function for the root of a tree:

\begin{equation*}
A(r) = (V(r) \frown \vec{g}, A(L(r)), A(R(r)))
\end{equation*}

\noindent where $\frown$ is the vector concatenation operator. Then:

\begin{equation*}
F^\prime(P) = \{ A(r) \mid r \in F(P)  \}
\end{equation*}

We refer to each entry of an augmented tree node's vector as a \emph{channel}. Next, we define tree convolution, an operation that maps a tree with $c_{in}$ channels to a tree with $c_{out}$ channels. The augmented forest is passed through a number of tree convolution layers. Details about tree convolution can be found in~\cite{tree_conv}. Here, we will provide a mathematical specification. Let a \emph{filterbank} be a matrix of size $3 \times c_{in} \times c_{out}$. We thus define the convolution of a root node $r$ of a tree with $c_{in}$ channels with a filterbank $f$, resulting in an structurally isomorphic tree with $c_{out}$ channels:

\begin{equation*}
(r * f) = (V(r) \frown L(r) \frown R(r) \times f, L(r) * f, R(r) * f)
\end{equation*}

We define the convolution of a forest of trees with a filterbank as the convolution of each tree in the forest with the filterbank. The output of the three consecutive tree convolution layers in the value network, with filterbanks $f_1$, $f_2$, and $f_3$, and thus be denoted as:

\begin{equation*}
T = ((F^\prime(P) * f_1) * f_2) * f_3
\end{equation*}

Let $final_{out}$ be the number of channels in $T$, the output of the consecutive tree convolution layers. Next, we apply a dynamic pooling layer~\cite{tree_conv}. This layer takes the element-wise maximum of each channel, flattening the forest into a single vector $W$ of size $final_{out}$. Dynamic pooling can be thought of as stacking each tree node's vectorized representation into a tall matrix of size $n \times final_{out}$, where $n$ is the total number of tree nodes, and then taking the maximum value in each matrix column.

Once produced, $T$ is passed through a final set of fully connected layers, until the final layer of the network produces a singular output. This singular output is used to predict the value of a particular state.

%%% Local Variables:
%%% mode: latex
%%% TeX-master: "main"
%%% End: